\documentclass[%
        twocolumn,
         longbibliography,
          amsmath,amssymb,
           aps,
            pra,
            ]{revtex4-1}
\usepackage{dcolumn}
\usepackage{bm}
\usepackage{graphicx,subfigure}
\graphicspath{{./figures/}}
\usepackage{xcolor}
\definecolor{sapphire}{HTML}{0067A5}
\usepackage[colorlinks=true,citecolor=sapphire,linkcolor=sapphire,urlcolor=sapphire]{hyperref}
\newcommand{\si}{\text{sin}}
\newcommand{\co}{\text{cos}}

\definecolor{smoke}{rgb}{0.65,0.65,0.65}

\begin{document}

\title{Rigidity transitions in zero-temperature polygons}
\author{M. C. Gandikota$^{1,2,*}$, Amanda Parker$^{1,3}$,  J. M. Schwarz$^{1,4}$}
\affiliation{$^1$Department of Physics and BioInspired Institute, Syracuse University, Syracuse, NY USA\\
    $^2$Department of Chemistry, Columbia University, New York, NY USA\\
$^3$SymBioSys, Chicago, Il USA\\ $^4$Indian
Creek Farm, Ithaca, NY USA}
\email{mcgandikota@gmail.com}
\date{\today}
\begin{abstract}
We study geometrical clues of a rigidity transition due to the emergence of a system-spanning state of self stress in under-constrained systems of individual polygons and spring networks constructed from such polygons. When a polygon with harmonic bond edges and an area spring constraint is subject to an expansive strain, we observe that convexity of the polygon is a necessary condition for such a self stress. We prove that the cyclic configuration of the polygon is a sufficient condition for the self stress.
This correspondence of geometry and rigidity is akin to the straightening of a one dimensional chain of springs to rigidify it. We predict the onset of the rigidity transition and estimate the transition strain using purely geometrical methods. These findings help determine the rigidity of an  area-preserving polygon just by knowing its geometry. Since two-dimensional spring networks can be considered as a network of polygons, we look for similar geometric features in  under-constrained spring networks under isotropic expansive strain. We observe that all polygons attain convexity at the rigidity transition such that the fraction of convex, but not cyclic, polygons predicts the onset of the rigidity transition. Acyclic polygons in the network correlate with larger tensions, forming effective force chains.  
\end{abstract}                                                                                   
\maketitle

\section{Introduction}
Athermal systems such as biopolymer networks can be modeled as disordered elastic networks~\cite{broedersz2014}.
    The linear elastic response of granular media modeled by frictionless soft sphere packings can also be represented by such disordered networks~\cite{alexander1998}. This is accomplished by mapping  every contact between neighboring spheres to the harmonic springs of a disordered spring network~\cite{wyart2005rigidity,wyart2008}.
Determining the rigidity of such disordered networks is a nontrivial problem that depends on both the topology and geometry of the network.

A network that is not rigid can be deformed while preserving the bond lengths. A \textit{floppy mode} is a deformation of the network that preserves the bond lengths to first order in the displacement of the vertices. While nontrivial floppy modes change the shape/geometry of the network, trivial floppy modes are just the global translations and rotations of the network. When all nontrivial floppy modes are removed, the network is said to be first-order, or infinitesimally, rigid. This is a microscopic definition of rigidity~\cite{connelly2015}. By adding bonds, or constraints, we change the topology of the network and rigidification is achieved when the number of degrees of freedom is equal to the number of independent constraints in the system~\cite{maxwell}. A canonical example of this phenomenon is that the infinitesimal rigidity of a two-dimensional spring network with central-force spring interactions can be determined solely by using a {\it combinatorial} theorem that identifies the number of independent constraints~\cite{laman}. In other words, no geometry needs to be invoked.

The other independent manner of rigidifying an under-constrained network is through a distribution of stresses in the network that adds up to zero net force on every vertex~\cite{lubensky2015}. Such a system spanning stress distribution is termed a state of self stress~\cite{alexander1998,connelly2015}. The simplest example of this is an under-constrained chain of springs which rigidifes when stretched out. The tension in the chain imposes a bending energy cost on all its transverse fluctuations~\cite{alexander1998}. Straining such under-constrained central-force spring networks can reposition the vertices such that a state of self stress can be established.

The difference between the number of nontrivial floppy modes $N_0$ and the number of states of self stress $N_S$ for a given configuration is determined by Maxwell-Calladine theorem as~\cite{calladine1978,lubensky2015},
\begin{equation}\label{maxwell}
N_0-N_S=dN-N_b-\frac{d(d+1)}{2},
\end{equation}
where $N_b$ is the number of constraints in the spring-network embedded in a $d$ dimensional space. This is a topological condition that does not involve any geometrical details of the network. However, the existence of a state of self stress is a question of geometry where we need to identify potential shapes that can satisfy force balance on the vertices.  

A canonical example for a state of self stress establishing rigidity in under-constrained networks is the case of two-dimensional disordered spring networks under isotropic expansive strain. Even though the order of the transition could not decidedly be found for spring networks derived from a diluted triangular lattice, a continuous phase transition in the bulk modulus appears to agree better with the numerical results~\cite{sheinman2012}. In contrast, for disordered spring networks derived from the contact network of jammed particles, strain in the form of isotropic expansion establishes a discontinuous rigidity transition in the bulk modulus~\cite{merkel2019}. Strain-induced rigidity transitions, such as these, are being  studied via several approaches~\cite{vermeulen2017,arzash2019,damavandi2021,damavandi2022,zhang2021,damavandi2022,lee2022}.

On the other hand, the statistical mechanics of shape (not rigidity) transitions in two-dimensional \textit{thermal} polymer rings with a pressure energy term has been carefully studied. These were studied as idealized models for membrane vesicles~\cite{rudnick1991,leibler1987,haleva2006,mitra2012}. Specifically, the polymer rings are constructed as $N$ number of springs and a pressure $p$ coupled with the enclosed area $A$. While the energetic pressure term $-pA$ favors inflation of the ring, entropic effects in these thermal models favor the crumpling of the ring~\footnote{There are more possible configurations for a crumpled polygon as compared to the inflated configurations.}. Thus, entropy counters the outward pressure force and serves as an `inward' pressure. In these shape-transition studies, the mean area $\left<A\right>$ is used as the order parameter and pressure is the tuning parameter. For self-intersecting rings, at the critical pressure $p_c\sim N^{-1}$, the ring blows up achieving infinite area at the transition point~\cite{rudnick1991,gaspari1993}. By replacing stretchable rods with rigid rods, a continuous phase transition between the crumpled and the inflated phases was found~\cite{haleva2006}. This behavior is different for self-avoiding rods where smooth crossovers between three distinct scaling regimes were recognized~\cite{leibler1987}.

While the rigidity transitions in zero temperature disordered spring networks under expansive strain are well studied, the geometry of the network i.e. the shapes of its constituent polygons which is crucial in sustaining states of self stress have not been given sufficient attention. In this paper, we attempt to bridge this gap by addressing some important shape aspects of this rigidity transition. To disassociate the constraints imposed by the network on the constituent polygons, we first study the correspondence of shape and rigidity in \textit{isolated} zero-temperature polygons under expansive strain. We then study the shapes of the polygons in the spring network at the point of transition in light of these findings.

The critical pressure of transition in a polygonal ring with a pressure energy term is $p_c=4\pi\; k_B T l^2/N$, where $l$ is the length of each edge of the loop. In the limit of zero temperature, the critical pressure for the polygon to blow up into a regular, cyclic polygon is simply $p_c=0^+$~\cite{haleva2006}.
However, the zero-temperature spring networks, which we are motivated by, have a rigidity transition at finite strain~\cite{wyart2008}.
To enable a nontrivial transition in athermal isolated polygons, we add  a quadratic energy of the form $k_AA^2$ with the constant $k_A>0$ to the Hamiltonian. By completing the square with the pressure energy $-pA$ as $k_A(A-p/(2k_A))^2$, the area-dependent energy term in the Hamiltonian would be $k_A(A-A_0)^2$ with $A_0=p/2k_A$.

The important results of our paper are as follows. We show that isolated polygons with an area conserving constraint have a rigidity transition under isotropic strain. We recognize that convexity of a polygon is a necessary condition  to sustain a state of self stress. We prove that a cyclic polygon - a polygon that can be circumscribed on a circle is a sufficient condition for the same.  
In contrast, when such polygons are used as building blocks to construct a disordered spring network, we find that it is not the cyclic polygons that mark the rigidity transition. The network is seen to be composed of both cyclic and acyclic polygons when it attains rigidity. Yet all polygons are strictly convex at the transition point.

\begin{figure}[t]
    \centering
    \includegraphics[width=0.45\textwidth]{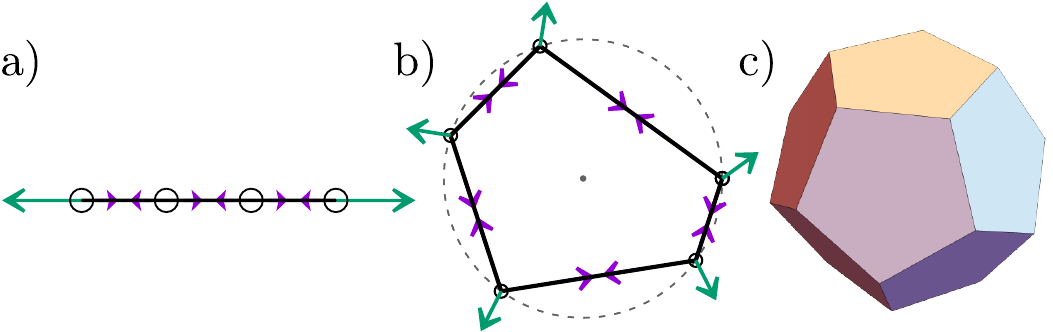}
    \caption{Correspondence of geometry and rigidity. a) A straightened out chain of springs rigidifies under extensional strain. b) A polygon under expansive strain is rigid when it attains a cyclic configuration. c) Convex polyhedrons are known to be rigid. While (a) and (b) are rigid due to the virtue of a system spanning state of self stress, (c) is rigid solely due to the shape-preserving constraints of the polyhedron faces.}\label{cauchy_intro}
\end{figure}

While a non-zero macroscopic elastic constant typically determines the rigidity of a polygon, we observe that we can also predict the rigidity of this under-constrained system using its geometry given its trivial topology. We know that a one-dimensional chain of edges when straightened out rigidifies via a state of self stress. The natural two-dimensional equivalent of this system - a polygon under expansive strain can sustain a system spanning state of self stress when it attains a cyclic configuration. The correspondence of shapes in judging the presence of a system-spanning state of self stress is reminiscent of an important result of Cauchy which guarantees the rigidity of three dimensional polyhedrons if the geometric condition of convexity is satisfied~\cite{cauchy1813,aigner2010}. While Cauchy's theorem does not apply to polygons, an extension of this theorem by Alexandrov proves that all convex polytopes in $\mathbb{R}^d$ with $d\geq3$ are rigid~\cite{pak2010lectures}.

The outline of the manuscript is as follows. We introduce the model in Sec. \ref{rplm} and present the numerical results in Sec. \ref{numerical}. We observe that there is no singularity associated with this transition in the thermodynamic limit. In Sec. \ref{necessary}, we show that convexity of the polygon is a necessary condition for a state of self stress. In Sec. \ref{cyclic_polygon}, we prove that the cyclic polygon is the unique configuration sustaining a state of self stress and is the geometrical signature of rigidity in the loop. In Sec. \ref{approach_R}, we identify the transition using a purely geometrical method by employing nontrivial floppy modes. In Sec. \ref{reg_pol}, we estimate the rigidity transition strain by approximating the initial random polygon as a regular polygon. In Sec. \ref{spring_sec}, we show numerical results that demonstrate that convex polygons which are \textit{acyclic} in general, mark the rigidity transition in periodic disordered spring networks.

\noindent \textit{Notation:} We represent vectors by lower case bold letters as in $\bm{r}$ and matrices by upper case bold letters as in $\bm{R}$.

\section{Random polygon model}\label{rplm}
To generate a random polygon with $N$ vertices ($N-$gon), we use Graham's algorithm \cite{rourke}. We randomly pick $N$ points in the $xy$ plane within a two dimensional box of dimensions $\widetilde{L}_B\times \widetilde{L}_B$. The nearest of these points to the `center of mass' of this set is chosen to be the first vertex of the polygon~\footnote{Graham's algorithm picks a point within the convex hull of these set of points. We are assuming here that the `center of mass' of the points is within the convex hull of the points. We perform this approximation for numerical simplicity.}. The rest of the points are sorted according to the angle of the vector joining the first vertex with each of the other $N-1$ points. This sorting assigns the set of points with vertex indices ${2,3,...,N}$. The polygon is constructed by joining these vertices in the order of the assigned indices. The area enclosed by this random polygon is $\widetilde{a}_0$ and the $i^{\text{th}}$ edge length is $\widetilde{l}_{i_0}$. While we construct an 8-gon constrained to a box of lengths $\widetilde{L}_B\times \widetilde{L}_B$, all $N$-gons with $N>8$, are constrained to a box of edge lengths $\sqrt{N/8}\;\widetilde{L}_B$. This ensures that the density of the random points in the plane does not scale with $N$.

Once we have the random polygon, the Hamiltonian of the system is defined as,
$$\widetilde{H}=\frac{1}{2}\widetilde{K}_\ell\sum_{i=1}^N(\widetilde{\ell}_i-\widetilde{\ell}_{i_0})^2+\frac{1}{2}\widetilde{K}_A(\widetilde{A}-\widetilde{A}_0)^2,$$
where $\widetilde{K}_\ell$ and $\widetilde{K}_A$ are the spring constants of two-body springs associated with the edges and the area spring, respectively. Variables $\widetilde{\ell}_i$ and $\widetilde{A}$ are the lengths of the $i^{\text{th}}$ spring and the area of the enclosed loop respectively while $\widetilde{\ell}_{i_0}$'s and $\widetilde{A}_0$ as the rest lengths of the two-body spring and the rest area of the ``area spring'', respectively. At zero strain, the rest lengths and rest area match those of the initial configuration of the random polygon such that $\widetilde{H}=0$. In this work, we measure lengths in units of $\widetilde{L}_B$ and energies in units of $\widetilde{K}_A\;\widetilde{L}_B^4$. In non-dimensionalized units, the Hamiltonian is,
\begin{equation}\label{1loop_hamiltonian}
H=\frac{1}{2}K_\ell\sum_{i=1}^N(\ell_i-\ell_{i_0})^2+\frac{1}{2}(A-A_0)^2,
\end{equation}
where $K_\ell=\widetilde{K}_\ell/\widetilde{K}_A\widetilde{L}_B^2$.
For a polygon embedded in $\mathbb{R}^2$ and $\hat{z}$ being the direction perpendicular to this plane, the area of the loop is taken to be the \textit{algebraic area} \cite{rudnick1991,gaspari1993,haleva2006},
\begin{equation}\label{shoelace}
A=\sum_{i=1}^N (\bm{r}_i\times\bm{r}_{i+1})\cdot\hat{z},
\end{equation}
where the position of the $i^{\text{th}}$ vertex is denoted by $\bm{r}_i$.
Note that we imply cyclicity of indices in $i+1$ and $N+1$ index is taken to be index 1.

    \begin{figure}[t]
        \centering
    \includegraphics[width=0.45\textwidth]{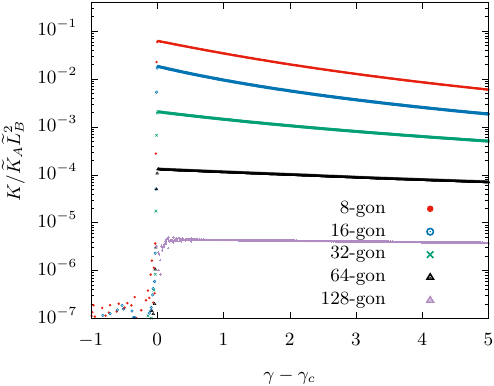}
    \caption[Expansive strain-induced rigidity transition in random polygons]{ A floppy to rigid transition observed for varying polygon sizes. Stiffness jumps at $\gamma=\gamma_c$, however this step-size decreases with polygon size. Every curve is averaged over 100 runs. }
        \label{oneloop_crossover}
    \end{figure}

\section{Numerical results}\label{numerical}
We can equivalently impose expansive strain on the system by decreasing $\ell_{i_0}$ of each spring proportionally or by increasing $A_0$ in Eq. (\ref{1loop_hamiltonian}), i.e., strain can be imposed by tuning down the dimensionless ratios $l_{i_0}/\sqrt{A_0}$. We choose to strain the polygon by increasing $A_0$. The strain $\gamma$ is defined as,
\begin{equation}\label{gamma}
    \gamma=\frac{A_0-a_0}{a_0},
\end{equation}
where $a_0$ is the initial area of the random polygon. Energy minimization for every imposed strain is performed numerically in C++ using the BFGS2 method in the \texttt{multimin} package of GNU scientific library~\cite{galassi2002gnu}. While Graham's algorithm~\cite{rourke} ensures a non-intersecting polygon at zero strain, we allow self-intersections of floppy polygons at non-zero strains. This retains the simplicity of the system for the purpose of energy minimization. We denote the minimized energy as $E$. Stress $\sigma$ and stiffness $K$ are defined as derivatives of the energy density as,
\begin{equation}
    \sigma=\frac{1}{A}\;\frac{dE}{d\gamma},
\end{equation}
and,
\begin{equation}
K=\frac{d\sigma}{d\gamma}.
\end{equation}

\begin{figure*}[t]
    \centering
    \includegraphics[width=0.95\textwidth]{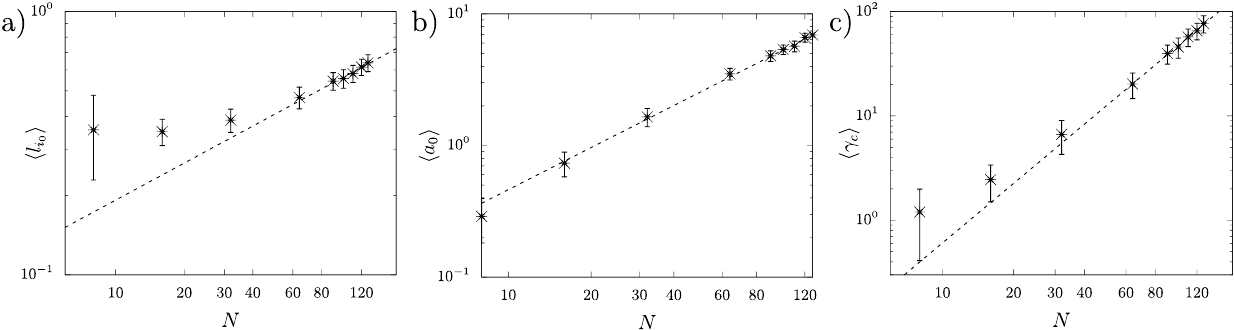}
    \caption{Finite-size scaling in random polygons.  
 a) The average edge length scales weakly as $N^{0.46}$. b) Average area of random polygon at zero strain scales as $N^{1.06}$.
c) Average critical strain of rigidity transition scales as $\sim N^{1.89}$ at large $N$. All scalings are obtained as fits to large $N$ limits of our simulations. The data points are obtained by averaging over 100 runs and error bars are standard deviation about the average.}
\label{ngammac}
\end{figure*}

For reporting numerical results, we choose the spring constants $K_\ell=1$. This choice does not change the critical transition strain.
The energy minimized configuration for regular polygons can be easily solved analytically due to the symmetry of the polygon (see Appendix. \ref{appendix_reg_pol}). For irregular polygons, we observe a floppy to rigid transition in the system. Stress is seen to be continuous across the transition point $\gamma_c$ and the stiffness jumps. The size of this jump decreases with $N$.
The finite size analysis for stiffness is reported in Fig. (\ref{oneloop_crossover}). Judging from the nature of the decrease in step-size of the stiffness with increasing $N$, there is no true phase transition, i.e., a transition where derivatives of the energy show singularities in the thermodynamic limit of $N\rightarrow\infty$. We, instead, have a smooth crossover between the floppy and rigid regimes.

Numerically we found that geometrical quantities such as average edge length $\left<l_{i_0}\right>$ weakly scales as $N^{0.46}$ and average initial area of random polygon $\left<a_0\right>$ scales as $N^{1.06}$. 
The critical strain scales with polygon size as $\left<\gamma_c\right>\sim N^{1.89}$ (see Fig. \ref{ngammac}). All the scalings are in the large $N$ limit of our simulations. 
In Sec. (\ref{reg_pol}), under a regular polygon approximation to predict critical strain at large $N$, we calculate the finite-size scaling of the critical strain using the scalings of geometrical quantities of the initial configurations - $\left<l_{i_0}\right>$ and $\left<a_0\right>$.

\section{Convexity is necessary for a system spanning state of self stress}\label{necessary}
    We now present two short arguments to show that convexity is necessary for the loop to sustain a state of self stress where non-zero forces due to the area spring and the tensions in the spring add up to a zero net force on each vertex of the polygon.

The force on the $i^{\text{th}}$ vertex due to the area spring is given by the negative gradient of the corresponding energy term in Eq. (\ref{1loop_hamiltonian}),
\begin{equation}\label{area force}
    \bm{f}_{A_i} = (A_0-A)\;\left(y_{i+1}-y_{i-1},x_{i-1}-x_{i+1}\right).
\end{equation}
This vector is perpendicular to the line joining the adjoining vertices of the $i^{\text{th}}$ vertex. This is easily seen by noticing that the dot product of this force with $\left(x_{i+1}-x_{i-1},y_{i+1}-y_{i-1}\right)$ is zero. Thus, the direction of force due to the area spring at any given vertex can be easily constructed geometrically. See (Fig. \ref{schem}).

\begin{figure}[t]
    \centering
    \includegraphics[width=0.35\textwidth]{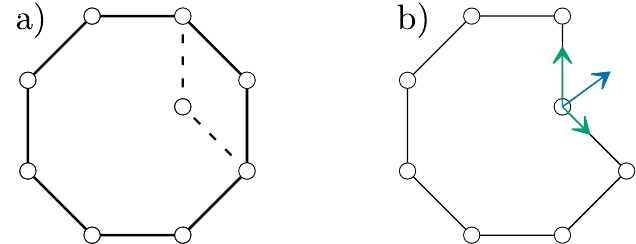}
    \caption[Convexity is necessary for rigidity]{Convexity is necessary for rigidity in random polygons with an area constraint. a) Upon an expansive strain, the non-convex configuration represented by dashed-line edges flip over to the solid-line edges. Thus the polygon achieves an increase in area without changing edge lengths. b) The pressure force of the area spring pointing outwards (blue arrow) and tensions in the two body springs (green arrows) cannot satisfy force balance at the mouth of the polygon.}\label{flip}
\end{figure}

An $i^{\text{th}}$ vertex is called an \textit{ear/mouth} if the line joining $i-1^{\text{th}}$ and $i+1^{\text{th}}$ vertices lie inside/outside the polygon~\cite{devadoss2011}. When the tuning parameter $A_0$ is increased (see Eq. \ref{1loop_hamiltonian}), a non-convex polygon attempts to increase its area without changing edge lengths. This is achieved by flipping the two dashed edges in Fig. (\ref{flip}a) into the two solid edges. A non-convex polygon can always increase its enclosed area without changing edge lengths via this discrete floppy mode motion which transforms a mouth to an ear
\footnote{If the increase in $A_0$ is small compared to the area enclosed by the rhombus formed by the dashed and solid lines, then the discrete flip of the edges in Fig. (\ref{flip}a) will be an over-compensation to the imposed strain and incurs an energy cost. Here, the response of the polygon will be less local. The nearest neighbor and the next-to-nearest neighbor edges of the mouth will continuously open to accommodate small increases in $A_0$.}.
Thus non-convex polygons cannot be rigid under expansive strain expansion.

 Another independent argument can be given as follows. When there is an outward force on the vertices due to the area spring, the edges should necessarily be in tension (as opposed to being in compression) to satisfy force-balance at all the ears.
Since a mouth is flanked by ears, this implies that the edges at the mouth are in tension. However, at the mouth, edges in tension cannot sustain a force-balance with the outward area spring force (See Fig. \ref{flip} b).
Since, force balance cannot be satisfied by non-convex loops, they cannot sustain a rigidifying state of self stress. The necessity of convexity is demonstrated in Fig. (\ref{ui} a) where convexity transition precedes the rigidity transition.

\section{Cyclicity is sufficient for a system spanning state of self stress}\label{cyclic_polygon}
In three dimensional spaces, Cauchy established an important connection between geometry and rigidity, assuming a trivial topology~\cite{cauchy1813,aigner2010}. First, the congruence theorem of Cauchy proves that two convex polyhedrons are congruent if corresponding faces of polyhedrons are congruent. A corollary of this theorem shows that a convex polyhedron with elastic faces is rigid i.e. the shape of the polyhedron cannot be changed without changing the shape of at least one of its faces. The geometry of the polyhedron is thus sufficient to judge the rigidity of the polyhedron. For the polygons with an area spring constrained to two dimensions, the transition point is again purely determined by the geometry of the polygon approaching the cyclic polygon. A \textit{cyclic polygon} is a polygon whose vertices can be inscribed on a circle which is referred to as the \textit{circumcircle}.  

For two-dimensional polygons, a fundamental result in Euclidean geometry shows that the maximum area of a flexible polygon is the polygon whose vertices lie on a common circle (Theorem 12.5a of \cite{niven1981}).  It is very intuitive to expect that when the loop cannot expand any more, it will become rigid. The subtlety here is displayed when we ask the question - how can the polygon shape which cannot increase area any more without changing edge lengths be the same shape that can support a state of self stress? We present analytical and numerical arguments  to show that the configuration of a cyclic polygon is a \textit{sufficient} condition to support a state of self stress.

\begin{figure}[t]
    \centering
    \includegraphics[width=0.3\textwidth]{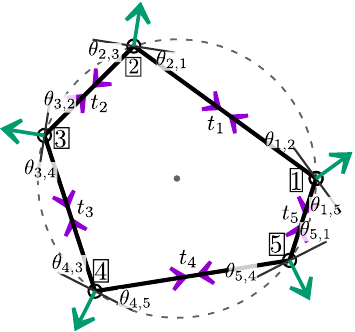}
    \caption{Schematic of a cyclic pentagon in a state of self stress. The black-dashed line is the circumcircle. Purple inward arrows on the edges indicate the inwards tension forces in the springs. Green outward arrows indicate the outward non-radial forces of the area spring. The thin solid black line segment at each vertex is the line perpendicular to the outward force of the area spring. $\theta_{i,j}$ is the angle between the `tangent' at vertex $i$ and the vertex $j$. Note that the term `tangent lines' we loosely use are not tangent to the circumcircle. }\label{schem}
    \end{figure}

A cyclic configuration of a $N$-gon reduces $2N$ degrees of freedom of $N$ vertices to $N+1$ degrees of freedom which are the $N$ number of angular coordinates of the vertices and the radius $r$. Without loss of generality we ignore the radius since it only serves as a scale factor and write the coordinates of the polygon vertices as,
\begin{equation}
        (x_i,y_i)=(\co\;\phi_i,\si\;\phi_i).
    \end{equation}
    The angles between the tangent at the $i^{\text{th}}$ vertex and the $j^\text{th}$ edge is denoted by $\theta_{i,j}$ as shown in Figure~\ref{schem}. We decompose the forces at each vertex along the direction of force of the area spring and the direction tangent to this. At each vertex, there are two force balance conditions along each of these directions. This system of simultaneous linear equations in the spring tensions $t_i$ and the forces due to the area spring $n_i$ can be neatly represented as a matrix equation $\bm{A}\bm{F}=\bm{0}$,

    \begin{widetext}
    \begin{equation}\label{geomat}
        \begin{bmatrix}
              \text{cos}\;\theta_{1,2}&0&0&\cdot\cdot\cdot&-\text{cos}\;\theta_{1,N}&0&0&\cdot\cdot\cdot&\cdot\cdot\cdot&0\\
              -\text{cos}\;\theta_{2,1}&\text{cos}\;\theta_{2,3}&0&\cdot\cdot\cdot&0&0&0&\cdot\cdot\cdot&\cdot\cdot\cdot&0\\
              \vdots&\vdots&\vdots&\vdots&\vdots&\vdots&\vdots&\vdots&\vdots\\
              \vdots&\vdots&\vdots&\vdots&\vdots&\vdots&\vdots&\vdots&\vdots\\
              0&0&\cdot\cdot\cdot&-\text{cos}\;\theta_{N,N-1}&\text{cos}\;\theta_{N,1}&0&0&\cdot\cdot\cdot&\cdot\cdot\cdot&0\\
              
              \text{sin}\;\theta_{1,2}&0&0&...&\text{sin}\;\theta_{1,N}&-1&0&0&\cdot\cdot\cdot&0\\
              \text{sin}\;\theta_{2,1}&\text{sin}\;\theta_{2,3}&0&0&\cdot\cdot\cdot&0&-1&\cdot\cdot\cdot&\cdot\cdot\cdot&0\\
              \vdots&\vdots&\vdots&\vdots&\vdots&\vdots&\vdots&\vdots&\vdots\\
              \vdots&\vdots&\vdots&\vdots&\vdots&\vdots&\vdots&\vdots&\vdots\\
              0&0&\cdot\cdot\cdot&\text{sin}\;\theta_{N,N-1}&\text{sin}\;\theta_{N,1}&0&0&\cdot\cdot\cdot&\cdot\cdot\cdot&-1
              \end{bmatrix}
              \begin{bmatrix}
              t_1\\
              t_2\\
          \vdots\\
          \vdots\\
              t_N\\
              n_1\\
              n_2\\
          \vdots\\
          \vdots\\
              n_N
              \end{bmatrix}
              =
              \begin{bmatrix}
                  0\\
                  0\\
          \vdots\\
          \vdots\\
                  0\\
                  0\\
                  0\\
          \vdots\\
          \vdots\\
                  0\\
              \end{bmatrix},
\end{equation}
\end{widetext}

\noindent where $\bm{A}$ is a purely geometrical $2N\times2N$ matrix and $F$ is the column matrix of forces. We show that the rank of the matrix $\text{Rank}(\bm{A})=2N-1$ by showing that the rank of the tangent components of the force, i.e., the first $N$ rows of matrix $\bm{A}$ is less than $N$.
For this purpose, let us consider matrix $\bm{B}$ constructed with the top $N$ rows and the first $N$ columns of matrix $\bm{A}$. We write the cosines of angles $\theta_{i,j}$ in the matrix as a dot product of unit tangent vector $\hat{p}_i$ and the tension unit vector $\hat{t}_{j}$ in the spring connecting the $i^{\text{th}}$ and the $j^{\text{th}}$ vertices. These unit vectors are defined as $\hat{p}_i=\bm{p}_i/p_i$ and $\hat{t}_i=\bm{t}_i/t_i$. Here,

\begin{equation}\label{pt}
\begin{split}
    \bm{t}_i&=(\text{cos}\phi_{i+1}-\text{cos}\phi_{i},\;\text{sin}\phi_{i+1}-\text{sin}\phi_{i}),\\
    \bm{p}_i&=(\text{cos}\phi_{i+1}-\text{cos}\phi_{i-1},\;\text{sin}\phi_{i+1}-\text{sin}\phi_{i-1}),
    \end{split}
    \end{equation}

\noindent and $t_i$, $p_i$ are their respective magnitudes.
In this section, $(i+1)^{\text{th}}$ index is considered as $(i+1)$ mod $N$.
We now decompose $\bm{B}$ as $\bm{P}^{-1}\bm{C}\bm{T}^{-1}$ where $\bm{P}^{-1}$ and $\bm{T}^{-1}$ are full rank diagonal matrices with elements $p_i^{-1}\delta_{i,j}$ and $t_i^{-1}\delta_{i,j}$ respectively. Since the rank of a matrix does not change on multiplication with a full rank matrix, $\text{Rank}(\bm{B})=\text{Rank}(\bm{C})$. The simplified matrix $\bm{C}$ for the tangential (Tangential directions are perpendicular to the force of the area spring at the vertex and are not tangential to the circumcircle) force balance at the vertices is then

\begin{equation}
        \begin{bmatrix}
              |\bm{p}_1\cdot\bm{t}_1|&0&0&...&-|\bm{p}_1\cdot\bm{t}_N|\\
              -|\bm{p}_2\cdot\bm{t}_1|&|\bm{p}_2\cdot\bm{t}_2|&0&...&0\\
              0&-|\bm{p}_3\cdot\bm{t}_2|&|\bm{p}_3\cdot\bm{t}_3|&...&0\\
              0&0&...&...&...\\
              0&0&...&-|\bm{p}_N\cdot\bm{t}_{N-1}|&|\bm{p}_N\cdot\bm{t}_N|\\
              \end{bmatrix}
    \end{equation}

    \noindent We take the modulus of the dot product since the angles between the tension vectors and the tangential vectors are always acute and, hence, the cosine of the angle must always be positive.

    The rank of a matrix is the dimension of the space spanned by its row or column vectors equivalently. To find the rank and nullity of $\bm{C}$, we attempt to find solutions (if they exist) to a set of $N$ number of equations given by

\begin{equation}
\sum_{i=1}^N \beta_i\;\bm{C}_i=\bm{0},
    \end{equation}

    \noindent where $\bm{C}_i$ is the $i^{\text{th}}$ row of matrix $\bm{C}$. While the trivial solution to $\beta_i$'s always exists, we seek nontrivial solutions. The coefficients $\beta_i$ can be solved for iteratively as

\begin{equation}
\beta_{i+1}=\beta_i\;\frac{|\bm{p}_i\cdot\bm{t}_i|}{|\bm{p}_{i+1}\cdot\bm{t}_{i}|}~~~~~~\text{for }i=1,2,...,N.
    \end{equation}

\noindent Using the coordinate representation in Eq. (\ref{pt}), and taking advantage of trigonometric identities, we obtain

\begin{equation}\label{betai}
    \begin{split}
        \beta_{i+1}=\beta_i\;\bigg|&\text{sin}\;\left(\frac{\phi_{i+1}-\phi_{i-1}}{2}\right)\text{cosec}\;\left(\frac{\phi_{i+2}-\phi_{i}}{2}\right)\\
    \times&\text{cos}\;\left(\frac{\phi_{i-1}-\phi_{i}}{2}\right)\text{sec}\;\left(\frac{\phi_{i+2}-\phi_{i+1}}{2}\right)\bigg|.\\
\end{split}
    \end{equation}

By iterating through the first $N-1$ equations, and cancellation of the (sine and cosine)/(cos and secant) terms of subsequent/next-subsequent iterations we can express $\beta_{N}$ in terms of $\beta_{1}$ as
\begin{equation}\label{betai}
    \begin{split}
        \beta_{N}=\beta_1\;\bigg|&\text{cosec}\;\left(\frac{\phi_{1}-\phi_{N-1}}{2}\right)\text{sin}\;\left(\frac{\phi_{2}-\phi_{N}}{2}\right)\\
    \times&\text{sec}\;\left(\frac{\phi_{N-1}-\phi_{N}}{2}\right)\text{cos}\;\left(\frac{\phi_{2}-\phi_{1}}{2}\right)\bigg|.\\
\end{split}
    \end{equation}

\noindent This relation is the same as the one implied by the $N^{\text{th}}$ equation of Eq. (\ref{betai}). Thus, the conditions on the $\beta_i$'s are not independent and a nontrivial solution to $\beta_i$'s can be written in terms of $\beta_1$ which serves as the chosen parameter for the set of solutions. The nullity of this matrix is one and the rank of the matrix is $N-1$. $\bm{C}$ not being full rank implies that $\bm{B}$ is not full rank. Since the rest of the row vectors in $\bm{A}$ are clearly independent among themselves and with the row vectors of $\bm{B}$ due to the $-1$ terms, rank of $\bm{A}$ is $2N-1$ and its nullity is one.

\begin{figure}[t]

    \centering
    \includegraphics[width=0.3\textwidth]{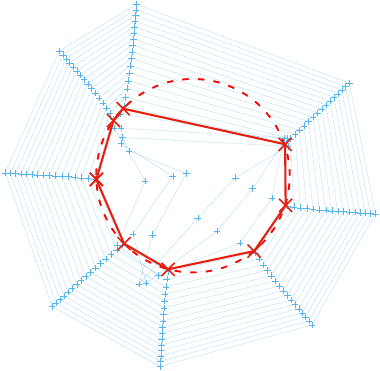}
    \caption{An irregular polygon under expansive strain rigidifies when a cyclic polygon configuration is achieved (shown in solid red line). The red dashed line is the circumcircle.}\label{octagon movie}
\end{figure}

While nullity of the geometric matrix $\bm{A}$ would remain zero in the floppy phase of the polygon allowing only trivial zero solutions for tensions in the polygon, post-rigidity transition, the nullity becomes one. This is akin to the emergence of a system spanning state of self stress under shear strain of spring networks~\cite{vermeulen2017}. Physically, for the random polygon, this means that for a given cyclic configuration of \textit{rigid} rods, at every imposed expansion strain on the loop, there exists a set of tensions parametrized by $t_1$ that satisfies force balance at all vertices. Note that not only at the critical strain, but at all strains post transition, the configuration of the polygon is cyclic for increasing radii.
The configurations of cyclic polygons are a sufficient condition to sustain a system-spanning state of self stress. These self stresses can then rigidity the loop. Given that the polygon which maximizes area for a given set of edges is unique~\cite{pinelis2005}, this further implies that at the rigidity transition, the cyclic polygon supports a \textit{unique} state of self stress. A unique state of self stress supporting the network's rigidity is seen in other strain-induced transitions as well~\cite{alexander1998amorphous, merkel2019, vermeulen2017}.

\section{Rigidity transition}
The critical strain of the rigidity transition in stretching a chain of springs or inflating random polygons is determined solely by the geometry of the systems. For a chain of floppy springs, the total contour length of the springs is sufficient to predict the critical strain of transition where the chain straightens out.  In this section, we discuss the approach to the transition and approximating the transition strain. The simplest method would be to calculate the asphericity of the polygon which should fall to zero at the point of transition. Alternatively, in Sec. (\ref{approach_R}) we delineate a method that explicitly shows the insufficiency of the nontrivial floppy modes in increasing the area of the polygon as it approaches the point of rigidity transition. In Sec. (\ref{reg_pol}), to approximate the transition strain in irregular polygons, we approximate the area of the cyclic irregular polygon at the transition point by the area of the corresponding regular polygon. Unlike the measurement of critical strain reported in Fig. (\ref{ngammac}), both these methods employ only the configurations of the polygons and do not require energy measurements to predict the rigidity transition.

\subsection{Identifying the approach to transition}\label{approach_R}
In this section, we predict the critical strain by observing the shape of the loop. At every strain, we use the energy minimized configuration to construct the rigidity matrix $\bm{R}$ for the two-body spring system whose zero modes are the floppy modes. We then evaluate the gradient of area $A$ expressed as a function of the nontrivial floppy variables of the polygon. The modulus of this gradient continuously decreases and falls to zero at the rigidity transition. This serves as a measure to indicate the transition.

For the purpose of formally defining the rigidity matrix $\bm{R}$, let $\bm{p}(\tau)$ be a continuous, analytic deformation of a network with $N$ vertices with $\tau$ denoting time such that $\bm{p}(0)=\bm{r}$ with
\begin{equation}
\bm{r} := (\bm{r}_1,\bm{r}_2,...\bm{r}_n).
\end{equation}
 The constraints of preserving the rest lengths of the two-body springs are
\begin{equation}\label{constraints}
    |\bm{r}_i-\bm{r}_j|^2=d^2_{ij},
\end{equation}
where $i,j$ are the vertices which flank the constraining edge of length $d_{ij}$, and $\bm{r}_i$ is the position vector of the $i^{\text{th}}$  vertex. Taking the derivative of Eq. (\ref{constraints}) with respect to $\tau$ at $\tau=0$, we have,
\begin{equation}\label{zer}
    (\bm{r}_i-\bm{r}_j)\cdot(\bm{p}_i^\prime-\bm{p}_j^\prime)=\bm{0}.
\end{equation}

\noindent $\bm{p}_i^\prime=\bm{p}_i^\prime(0)$ is the velocity of the $i^{\text{th}}$ vertex which deforms the network infinitesimally. Instead of preserving the constraints exactly as in Eq. (\ref{constraints}), the above set of equations preserve  the constraints only to first order in displacements. The system of equations that represents this first-order theory may be written in terms of a matrix equation
\begin{equation}\label{I order}
    \bm{R}(\bm{r})\bm{p}^\prime=\bm{0},
\end{equation}
where $\bm{R}(\bm{r})$ is the rigidity matrix. A network is first-order rigid if there are no nontrivial solutions to Eq. (\ref{I order}) with the trivial solutions being rigid body displacements of the entire network. A nonzero solution to this equation is a nontrivial floppy mode. Floppy modes are vectors that span the null space of the transformation. The motion of the bonds in a floppy mode is such that the velocity of each directed bond $\bm{p}^\prime_i-\bm{p}^\prime_j$ is perpendicular to the direction of the bond $\bm{r}_i-\bm{r}_j$, thus satisfying Eq. (\ref{zer}).

When expansive strain is imposed on the polygon by changing the tuning parameters, the polygon responds by moving to a configuration that minimizes the elastic energy (see Eq. \ref{1loop_hamiltonian}). Until the rigidity transition, this can be achieved via floppy motion: motion that preserves the lengths of the springs while also keeping the area constant.
For $N_S=0$ and $d=2$, we know from Eq. (\ref{maxwell}) that the number of floppy modes $N_0$ is $N-3$. We seek to write the change in position vector of the network $\Delta \bm{r}$ in terms of $N_0$ number of independent floppy variables,
\begin{equation}
\bm{u} := (u_1,u_2,...u_{N_0}).
\end{equation}

\begin{figure*}[t]
    \centering
    \includegraphics[width=0.75\textwidth]{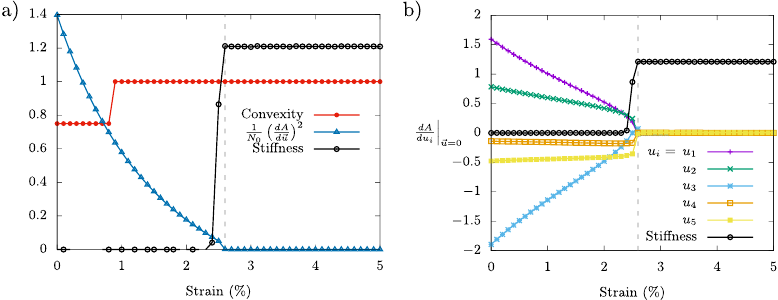}
    \caption{The approach to the rigidity transition can be followed by measuring the capacity of the nontrivial floppy modes to increase the area of the polygon. The irregular octagon used here has five nontrivial floppy modes. a) Convexity being a necessary condition for rigidity, the convexity transition precedes the rigidity transition. The rigidity transition coincides with the normalized gradient square of area, $1/N_0(dA/d\bm{u})^2$ approaching zero. b) All the five components of the gradient of the area goes to zero at the point of transition.}
    \label{ui}
\end{figure*}

We choose independent definitions for nontrivial floppy variables which we encode in the matrix equation $\bm{F}\Delta \bm{r}=\bm{u}$. Three more constraining equations are provided by fixing the two translations and rotation of the system. The associated matrix equation is denoted by $\bm{G}\Delta\bm{r}=\bm{0}$.
Finally Eq. (\ref{I order}) provides a set of $N$ equations as $\bm{R}(\bm{r})\Delta\bm{r} = \bm{0}$.

\begin{table}[h!]
    \centering
\begin{tabular}{c|c}
Dimensions & Matrix equations\\
\hline
$N_b\times 2N$& $\bm{R}\; \Delta \bm{r}=\bm{0}$\\
$3\times 2N$& $\bm{G}\; \Delta \bm{r} =\bm{0}$\\
$N_0\times 2N$& $\bm{F}\; \Delta \bm{r} = \bm{u}$\\
\hline
    $2N\times 2N$& $\bm{R}^\prime \; \Delta \bm{r} = \bm{u^\prime}$,
\end{tabular}
\end{table}

In the last row, we have vertically stacked these equations. In block notation of matrices the same operation may be written as,
    \begin{equation}
        \bm{R}^\prime=
        \begin{bmatrix}
\bm{R}\\
\bm{G}\\
\bm{F}\\
        \end{bmatrix}
        ,~~~
        \bm{u}^\prime=
        \begin{bmatrix}
\bm{0}\\
\bm{0}\\
\bm{u}\\
        \end{bmatrix}
        .
    \end{equation}

    \noindent Note that the dimensions of the zero vectors differ.
Assuming its invertibility, we invert the square matrix $\bm{R}^\prime$ to express the coordinate displacements of the system in terms of the nontrivial floppy variables,

$$\Delta \bm{r}(\bm{u})=(\bm{R}^\prime)^{-1} \bm{u^\prime}.$$

On imposition of strain, the displacements of vertices which obey all the spring constraints to first-order and that exclude trivial translations and rotation is a function of the independent non trivial floppy variables.

The area of the polygon can be found using Green's theorem/shoelace formula, Eq. (\ref{shoelace}). Upon strain imposition, we have,
\begin{equation}
\bm{r}^\prime(\bm{u})=\bm{r}+\Delta\bm{r}(\bm{u}).
\end{equation}
The geometric measure for the rigidity transition is,
\begin{equation}
    \bm{\nabla} A(\bm{r}^\prime)\bigg|_{\bm{u}=\bm{0}}=\left.\frac{dA(\bm{r}+\Delta \bm{r}(\bm{u}))}{d\bm{u}}\right\vert_{\bm{u}=\bm{0}}=\bm{0},
\end{equation}
i.e. the gradient of the area function with respect to the independent nontrivial floppy variables is uniformly zero. If an expansive area strain is imposed at this configuration, it cannot be achieved via the motion of nontrivial floppy variables and an energy cost needs to be paid.

Inversion of the rigidity matrix was done using the \texttt{SymPy} package of python~\cite{meurer2017sympy}. As an example, we use a two dimensional irregular octagon which has $N_0=5$ number of floppy modes (see Eq. (\ref{maxwell}) with $N_S=0$). The choice of nontrivial floppy variables $\bm{u}$ is,

\begin{equation}
\begin{split}
    u_1&=\Delta x_6+ \Delta y_6,\\
    u_2&=\Delta x_7+ \Delta y_5,\\
    u_3&=\Delta x_5+ \Delta y_7,\\
    u_4&=\Delta x_8+ \Delta y_4,\\
    u_5&=\Delta x_4+ \Delta y_8.
    \end{split}
    \end{equation}
The results are reported in Fig. (\ref{ui}). We see that the gradient of the area goes to zero at the critical strain.
The gradient of the area function approaching zero identifies the critical strain in a similar fashion as the singular value of the equilibrium matrix approaching zero identifies the rigidity transition in a shear strained network~\cite{vermeulen2017}.
We also conclude from the loss of all nontrivial zero modes at the transition that the system is microscopically rigid.

\subsection{Approximation of transition strain}\label{reg_pol}

\begin{figure}[t]
    \centering
    \includegraphics[width=0.45\textwidth]{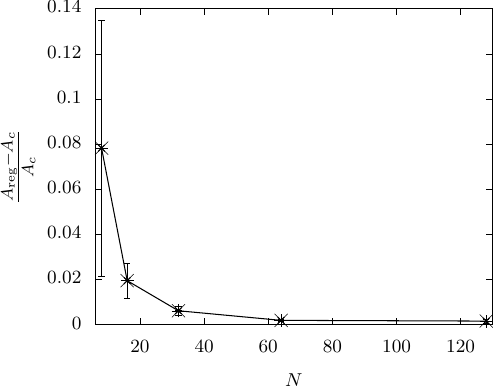}
    \caption{Regular polygons approximately predict critical strain of irregular polygons. $A_c$ is the numerically determined area of a cyclic polygon at critical strain and $A_\text{reg}$ is the area of the approximated regular polygon. The approximation is seen to work well in the $N\rightarrow\infty$ limit. Error bars are standard deviations about the average of 100 runs. The solid line is a guide to the eye. }\label{reg}
\end{figure}

To predict the transition strain, we need to know the area of the cyclic polygon corresponding to the initial edge length distribution of the polygon.
Finding the area of general cyclic polygons given the edges is an area of ongoing research in mathematics. For $N=4$, we can find the area of irregular cyclic polygons using Brahmagupta's formula,
\begin{equation}
\text{Area}_{\text{ cyclic quadrilateral}} = \sqrt{(s-a)(s-b)(s-c)(s-d)},
\end{equation}
where $a,b,c,d$ are the lengths of edges and $s$ is the semi-perimeter. The area of cyclic pentagons and hexagons have been found in more recent work~\cite{robbins1994,robbins1995}. It is a nontrivial problem to find an exact expression for the maximum area of the polygon. However, the area of any regular polygon (Polygons with equal edge lengths) is easily expressible in terms of its edge length. Here, we make a simplifying approximation by considering an irregular polygon as a regular polygon whose edge length $l_0$ is the average edge length of the irregular polygon. The area of this regular polygon is denoted as $A_{\text{reg}}$.
\begin{equation}
A_{\text{reg}}=\frac{N}{4}\;l_0^2\;\text{cot}(\psi/2),
    \end{equation}
    where $\psi$ is the angle subtended by the edges at the center of the polygon.
This approximation holds well in the limit of $N\rightarrow\infty$ (see Fig. \ref{reg}) simply because the area of both irregular polygons and regular polygons approach the area of the circumcircle with radius $R$ which satisfies the implicit equation,
\begin{equation}\label{cyclic_R}
    \sum_{i=1}^N \text{sin}^{-1} \left(\frac{\ell_{i_0}}{2R} \right) = \pi
\end{equation}
In the limit of large $N$, we have, $l_{i_0}/2R\rightarrow0$. The above implicit equation, under this limit, simplifies to $\sum_{i=1}^N l_{i_0}=2\pi R$, which is the perimeter of a circle of radius $R$.

 Using $\psi\sim N^{-1}$ and the numerically calculated finite-size scaling of $\left<l_{i_0}\right>$ and $\left<a_0\right>$ in Sec. (\ref{numerical}), we can predict the scaling of $\left<\gamma_c\right>$ in the large $N$ limit by substituting $A_c$ as $A_{reg}$ in Eq. (\ref{gamma}). We find that under this approximation $\left<\gamma_c\right>$ should scale as $N^{1.86}$ which agrees well with the numerical exponent which is 1.89.

\begin{figure*}[t]
    \centering
    \includegraphics[width=0.95\textwidth]{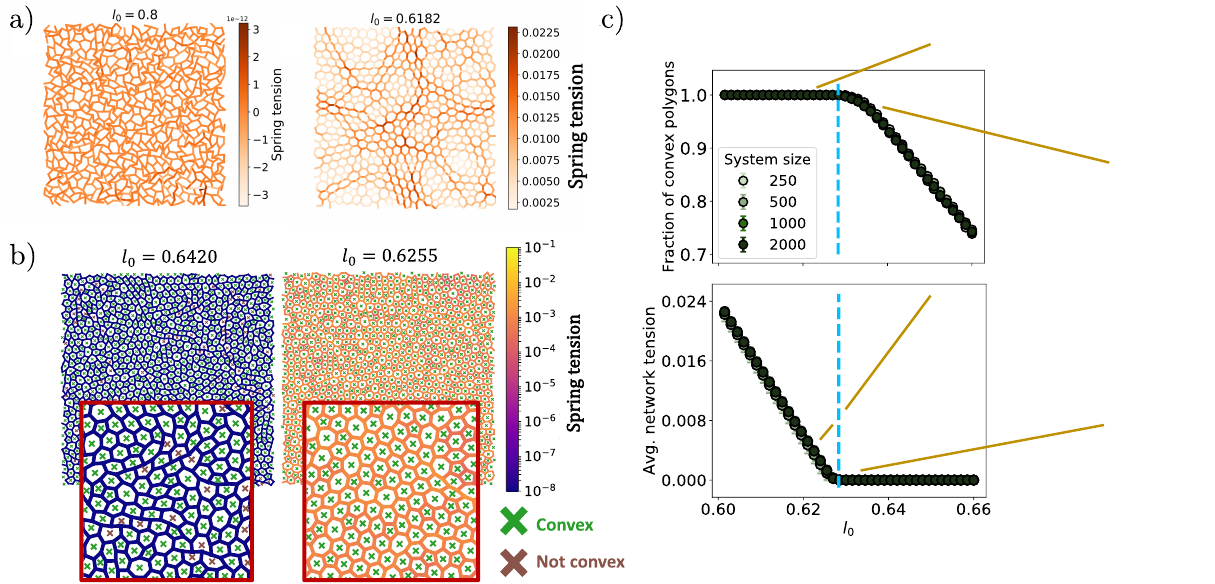}
    \caption{Isotropic strain-induced rigidity transition in under-constrained periodic spring networks. a) The spring network is floppy at $l_0=0.8$ and strain-rigidified at $l_0=0.6182$. Note that lower the $l_0$, more is the strain. b) Left: A floppy network does not have any spring tensions and constituent loops in the network need not be convex. Right: A rigidified network has non-zero spring tensions and all polygons are convex. Note the scale bar difference in color map of tensions in springs pre/post transition. c) Top/Bottom: Convexity/Rigidity transition as a function of isotropic strain. Convexity transition in the spring network coincides with the rigidity transition. The blue dashed line marks the point of rigidity transition. The blow up plots at the transition show weak finite size effects.}\label{amanda}
    \end{figure*}
 
\section{Spring networks}\label{spring_sec}
For random polygons, we established that convexity is necessary for a system spanning state of self stress in Sec. \ref{necessary} and that cyclicity is sufficient to guarantee a system spanning state of self stress in Sec. \ref{cyclic_polygon}. How do such geometric signatures translate to a system that is built out of random polygons? At the rigidity transition, do all polygons become cyclic polygons? These are the two main questions we address in this section.

We consider a two-dimensional spring network with the energy
\begin{equation}
H_{SN}=\sum_{i=1}^N(\ell_i-\ell_0)^2,
    \end{equation}
where $l_i$ is the spring length of the $i^{\text{th}}$ spring and $l_0$ is their rest length.  The network is three-fold coordinated and, therefore, under-constrained via Maxwell constraint counting. We implement periodic boundary conditions in both directions such that the network is constrained to a torus.  With this construction, we no longer require the area spring constraint. An isotropic strain is then imposed by decreasing the rest length $l_0$. Note that this choice is different from the one made in~\cite{merkel2019} where isotropic strain is imposed by proportionally decreasing the rest lengths $l_{i_0}$ of all springs.

As $l_0$ is decreased, the energy of the system is minimized using FIRE minimization for a modified version of cellGPU for two-body springs~\cite{sussman2017}. Once the average tension in the spring network is non-zero, we characterize it as rigid.  In our simulations (see Fig. \ref{amanda} b), we see that at the rigidity transition, all polygons defined by the bonds connecting vertices become convex. We also see in Fig. \ref{amanda} c that the convexity transition in the spring network - the strain at which all the loops in the network become convex - is the same as the point of rigidity transition. Additionally, the approach to transition can be judged by measuring the fraction of convex polygons since this fraction increases linearly with strain (see Fig. \ref{amanda} c). As for cyclicity, Fig. (\ref{amanda} a) clearly shows that post-transition strain, all polygons are {\it not} cyclic. Interestingly, the polygons supporting above-average tensions in the network appear less cyclic than the polygons supporting less than average tensions.

The important geometric signature is not cyclicity but convexity.
This is particularly interesting when contrasted against an isolated polygon where convexity \textit{precedes} the rigidity transition (see Fig. \ref{ui}a).
Interestingly, prior work studying ordered spring networks consisting of polygons of equal edge lengths with periodic boundary conditions demonstrates that all polygons are cyclic at the rigidity transition~\cite{guiducci2015}. In the presence of disorder, not all polygons are equal in terms of their cyclicity. As loops of polygons form, those polygons composed of higher-tension springs are more stressed in a particular direction.  Being more stressed, or tensioned in a particular direction, causes them to be acyclic.  However, the necessary condition of convexity still prevails, even in the disordered network. With convexity as the geometric signature, perhaps understanding of the onset of rigidity may be enhanced with a focus on loops of polygons as opposed to focus on stretching out lines of springs, which does not, in of itself, relate directly to convexity. Coincidentally, convex and nonconvex loops can map to a Boolean variable such that a mapping to an Ising-like description is possible, provided one quantifies the interaction between convex polygons, etc. One can also imagine that including area-conservation constraints for each polygon may modify the interaction such that an Ising spin glass-like description is possible given the additional constraints. It would indeed be interesting to pursue this line of inquiry further.

\section{Discussion}\label{discussion}

In the context of rigidity theory, Connelly proved that for a polygon, there exists an expanding motion that conserves the edge lengths. Such a motion convexifies the polygon and increases its area~\cite{connelly2000straightening}. The proof involves the addition of struts~\footnote{Struts are elastic rods which are in compression and hence want to expand.} connecting the edges in the polygon. This is reminiscent of a `pressure' within the polygon. In another work, such an expansive motion inspired by electrostatic charges has also been studied~\cite{cantarella2004}. In contrast to this, our work does not incorporate struts.

For a random polygon made up of harmonic springs and including an additional  area spring, we demonstrate that while a convex configuration is necessary, the cyclic configuration is sufficient for sustaining a system spanning state of self stress. Such a self stress can then rigidify the polygon with expansion strain. An isolated loop attaining a cyclic polygon configuration at the critical strain of rigidity transition, is independent of the details of the nonlinearities in the Hamiltonian. While we have explicitly proved that the cyclic configuration of the polygon can sustain a state of self stress, our numerical simulations show that such a self stress rigidifies the polygon as evidenced in the non-zero stiffness moduli. We, therefore, have arrived at a two-dimensional version of Cauchy's geometry-rigidity correspondence for isolated structures that are topologically trivial. It would be interesting to study the nature of this transition for a thermal system with a quadratic area-energy term and investigate the existence of a phase transition as previously done in the context of polymer rings with pressure coupled to area~\cite{rudnick1991,leibler1987,haleva2006,mitra2012}.  

Given our result for an individual polygon, a natural extension would be to investigate the geometry-rigidity correspondence for two polygons sharing an edge. There, we anticipate that the cyclicity-rigidity correspondence for each polygon still holds.  One can then naturally extend the investigation further by considering multiple polygons with shared edges to form a network of polygons formed with harmonic springs to arrive at a spring network. We numerically study this particular limit by imposing periodic boundary conditions. The network which now rests on a toroidal surface does not need area springs to sustain a state of self stress. We find that while the cyclic configuration is important for the case of the isolated random polygon to attain rigidity, all polygons are not cyclic for a strain-rigidified, under-constrained spring-network. However, the data suggests that all polygons are convex at the transition. Intriguingly, earlier work on an ordered, under-constrained spring network reports that the network rigidity correlates with the cyclicity of the polygons making up the network~\cite{guiducci2015}. How such a correlation is modified for an infinitesimal amount of disorder is an open question.

For the disordered, under-constrained spring networks studied here, it would be useful to study the correlation of the cyclicity of the constituent polygons in the network and the average tensions in such a loop. We expect the two quantities to be anti-correlated. More specifically, polygons that are less cyclic may become part of the `force chains' in the network that sustain the system spanning state of self stress. Of course, force chains have long been studied in granular systems~\cite{Mueth1998,Majmudar2005}.  Moreover, a correspondence between geometry and rigidity has been established in force tilings of particle packings to, again, highlight the importance of the correspondence~\cite{Sarkar2013}.  Specifically, for frictionless particle packings, all force tilings are convex in the rigid phase~\cite{Sarkar2013}. As we work to understand the correspondence between geometry and rigidity in general disordered systems, hopefully we can adopt a ``rigid or not by looking'' approach, at least in some limits.

\section*{Acknowledgments}
MCG acknowledges useful discussions with Manu Mannattil.
JMS acknowledges financial support from NSF-DMR-1832002 and an Isaac Newton Award for Transformative Ideas during the COVID19 Pandemic from the DoD.

\section{Appendix: Regular polygons}\label{appendix_reg_pol}
\renewcommand{\theequation}{A\thechapter.\arabic{equation}}

\begin{figure}[t]
    \centering
    \includegraphics[width=0.45\textwidth]{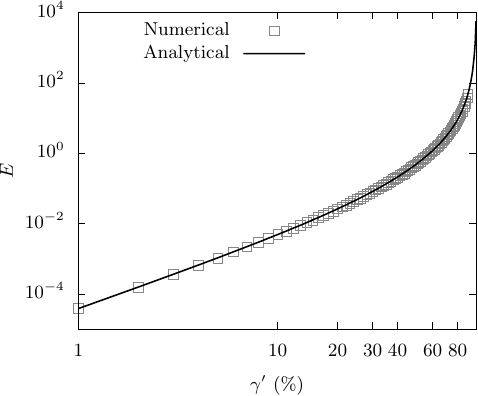}
    \caption{ Energy of regular polygon under isotropic expansive strain imposed via the area spring. Analytical solution (solid-line) is possible due to the symmetry of regular polygons. Note that we use $\gamma^\prime$ strain where $A_0$ spans very large values. }
    \label{reg_polygon}
\end{figure}

For a regular polygon, we can analytically solve for the minimized energy configuration under expansive strain. A regular polygon has only one degree of freedom - the edge length $l$. The area of the regular polygon is $N\;l^2/(4\;\text{tan}(\pi/N))$. An exact expression can be found by minimizing Eq. (\ref{1loop_hamiltonian}) with respect to $l$ at each strain. This gives a cubic equation in $l$ which was solved in \texttt{MATHEMATICA}. Instead of using, strain $\gamma$ as defined in Eq. (\ref{gamma}), we increase $A_0$ as $A_0=a_0/(1-\gamma^{\prime\;2})$ with $0\leq\gamma^\prime<1$ to capture the complete range of possible $A_0$'s. See Fig. (\ref{reg_polygon}).

\bibliography{rigidity_transitions}

\begin{thebibliography}{45}%
\makeatletter
\providecommand \@ifxundefined [1]{%
 \@ifx{#1\undefined}
}%
\providecommand \@ifnum [1]{%
 \ifnum #1\expandafter \@firstoftwo
 \else \expandafter \@secondoftwo
 \fi
}%
\providecommand \@ifx [1]{%
 \ifx #1\expandafter \@firstoftwo
 \else \expandafter \@secondoftwo
 \fi
}%
\providecommand \natexlab [1]{#1}%
\providecommand \enquote  [1]{``#1''}%
\providecommand \bibnamefont  [1]{#1}%
\providecommand \bibfnamefont [1]{#1}%
\providecommand \citenamefont [1]{#1}%
\providecommand \href@noop [0]{\@secondoftwo}%
\providecommand \href [0]{\begingroup \@sanitize@url \@href}%
\providecommand \@href[1]{\@@startlink{#1}\@@href}%
\providecommand \@@href[1]{\endgroup#1\@@endlink}%
\providecommand \@sanitize@url [0]{\catcode `\\12\catcode `\$12\catcode
  `\&12\catcode `\#12\catcode `\^12\catcode `\_12\catcode `\%12\relax}%
\providecommand \@@startlink[1]{}%
\providecommand \@@endlink[0]{}%
\providecommand \url  [0]{\begingroup\@sanitize@url \@url }%
\providecommand \@url [1]{\endgroup\@href {#1}{\urlprefix }}%
\providecommand \urlprefix  [0]{URL }%
\providecommand \Eprint [0]{\href }%
\providecommand \doibase [0]{http://dx.doi.org/}%
\providecommand \selectlanguage [0]{\@gobble}%
\providecommand \bibinfo  [0]{\@secondoftwo}%
\providecommand \bibfield  [0]{\@secondoftwo}%
\providecommand \translation [1]{[#1]}%
\providecommand \BibitemOpen [0]{}%
\providecommand \bibitemStop [0]{}%
\providecommand \bibitemNoStop [0]{.\EOS\space}%
\providecommand \EOS [0]{\spacefactor3000\relax}%
\providecommand \BibitemShut  [1]{\csname bibitem#1\endcsname}%
\let\auto@bib@innerbib\@empty
\bibitem [{\citenamefont {Broedersz}\ and\ \citenamefont
  {MacKintosh}(2014)}]{broedersz2014}%
  \BibitemOpen
  \bibfield  {author} {\bibinfo {author} {\bibfnamefont {Chase~P}\ \bibnamefont
  {Broedersz}}\ and\ \bibinfo {author} {\bibfnamefont {Fred~C}\ \bibnamefont
  {MacKintosh}},\ }\bibfield  {title} {\enquote {\bibinfo {title} {Modeling
  semiflexible polymer networks},}\ }\href@noop {} {\bibfield  {journal}
  {\bibinfo  {journal} {Reviews of Modern Physics}\ }\textbf {\bibinfo {volume}
  {86}},\ \bibinfo {pages} {995} (\bibinfo {year} {2014})}\BibitemShut
  {NoStop}%
\bibitem [{\citenamefont {Alexander}(1998{\natexlab{a}})}]{alexander1998}%
  \BibitemOpen
  \bibfield  {author} {\bibinfo {author} {\bibfnamefont {Shlomo}\ \bibnamefont
  {Alexander}},\ }\bibfield  {title} {\enquote {\bibinfo {title} {Amorphous
  solids: their structure, lattice dynamics and elasticity},}\ }\href@noop {}
  {\bibfield  {journal} {\bibinfo  {journal} {Physics reports}\ }\textbf
  {\bibinfo {volume} {296}},\ \bibinfo {pages} {65--236} (\bibinfo {year}
  {1998}{\natexlab{a}})}\BibitemShut {NoStop}%
\bibitem [{\citenamefont {Wyart}(2005)}]{wyart2005rigidity}%
  \BibitemOpen
  \bibfield  {author} {\bibinfo {author} {\bibfnamefont {M}~\bibnamefont
  {Wyart}},\ }\bibfield  {title} {\enquote {\bibinfo {title} {On the rigidity
  of amorphous solids},}\ }in\ \href@noop {} {\emph {\bibinfo {booktitle}
  {Annales de Physique}}},\ Vol.~\bibinfo {volume} {30}\ (\bibinfo
  {organization} {Springer Verlag},\ \bibinfo {year} {2005})\ pp.\ \bibinfo
  {pages} {1--96}\BibitemShut {NoStop}%
\bibitem [{\citenamefont {Wyart}\ \emph {et~al.}(2008)\citenamefont {Wyart},
  \citenamefont {Liang}, \citenamefont {Kabla},\ and\ \citenamefont
  {Mahadevan}}]{wyart2008}%
  \BibitemOpen
  \bibfield  {author} {\bibinfo {author} {\bibfnamefont {M}~\bibnamefont
  {Wyart}}, \bibinfo {author} {\bibfnamefont {H}~\bibnamefont {Liang}},
  \bibinfo {author} {\bibfnamefont {A}~\bibnamefont {Kabla}}, \ and\ \bibinfo
  {author} {\bibfnamefont {L}~\bibnamefont {Mahadevan}},\ }\bibfield  {title}
  {\enquote {\bibinfo {title} {Elasticity of floppy and stiff random
  networks},}\ }\href@noop {} {\bibfield  {journal} {\bibinfo  {journal}
  {Physical review letters}\ }\textbf {\bibinfo {volume} {101}},\ \bibinfo
  {pages} {215501} (\bibinfo {year} {2008})}\BibitemShut {NoStop}%
\bibitem [{\citenamefont {Connelly}\ and\ \citenamefont
  {Guest}(2022)}]{connelly2015}%
  \BibitemOpen
  \bibfield  {author} {\bibinfo {author} {\bibfnamefont {Robert}\ \bibnamefont
  {Connelly}}\ and\ \bibinfo {author} {\bibfnamefont {Simon~D.}\ \bibnamefont
  {Guest}},\ }\href@noop {} {\emph {\bibinfo {title} {Frameworks, Tensegrities,
  and Symmetry}}}\ (\bibinfo  {publisher} {Cambridge University Press},\
  \bibinfo {year} {2022})\BibitemShut {NoStop}%
\bibitem [{\citenamefont {Maxwell}(1864)}]{maxwell}%
  \BibitemOpen
  \bibfield  {author} {\bibinfo {author} {\bibfnamefont {J~Clerk}\ \bibnamefont
  {Maxwell}},\ }\bibfield  {title} {\enquote {\bibinfo {title} {L. on the
  calculation of the equilibrium and stiffness of frames},}\ }\href@noop {}
  {\bibfield  {journal} {\bibinfo  {journal} {The London, Edinburgh, and Dublin
  Philosophical Magazine and Journal of Science}\ }\textbf {\bibinfo {volume}
  {27}},\ \bibinfo {pages} {294--299} (\bibinfo {year} {1864})}\BibitemShut
  {NoStop}%
\bibitem [{\citenamefont {Laman}(1970)}]{laman}%
  \BibitemOpen
  \bibfield  {author} {\bibinfo {author} {\bibfnamefont {G.}~\bibnamefont
  {Laman}},\ }\bibfield  {title} {\enquote {\bibinfo {title} {On graphs and
  rigidity of plane skeletal structures},}\ }\href@noop {} {\bibfield
  {journal} {\bibinfo  {journal} {J. Engineering Math.}\ }\textbf {\bibinfo
  {volume} {4}},\ \bibinfo {pages} {331--340} (\bibinfo {year}
  {1970})}\BibitemShut {NoStop}%
\bibitem [{\citenamefont {Lubensky}\ \emph {et~al.}(2015)\citenamefont
  {Lubensky}, \citenamefont {Kane}, \citenamefont {Mao}, \citenamefont
  {Souslov},\ and\ \citenamefont {Sun}}]{lubensky2015}%
  \BibitemOpen
  \bibfield  {author} {\bibinfo {author} {\bibfnamefont {TC}~\bibnamefont
  {Lubensky}}, \bibinfo {author} {\bibfnamefont {CL}~\bibnamefont {Kane}},
  \bibinfo {author} {\bibfnamefont {Xiaoming}\ \bibnamefont {Mao}}, \bibinfo
  {author} {\bibfnamefont {Anton}\ \bibnamefont {Souslov}}, \ and\ \bibinfo
  {author} {\bibfnamefont {Kai}\ \bibnamefont {Sun}},\ }\bibfield  {title}
  {\enquote {\bibinfo {title} {Phonons and elasticity in critically coordinated
  lattices},}\ }\href@noop {} {\bibfield  {journal} {\bibinfo  {journal}
  {Reports on Progress in Physics}\ }\textbf {\bibinfo {volume} {78}},\
  \bibinfo {pages} {073901} (\bibinfo {year} {2015})}\BibitemShut {NoStop}%
\bibitem [{\citenamefont {Calladine}(1978)}]{calladine1978}%
  \BibitemOpen
  \bibfield  {author} {\bibinfo {author} {\bibfnamefont {Christopher~R}\
  \bibnamefont {Calladine}},\ }\bibfield  {title} {\enquote {\bibinfo {title}
  {Buckminster fuller's “tensegrity” structures and clerk maxwell's rules
  for the construction of stiff frames},}\ }\href@noop {} {\bibfield  {journal}
  {\bibinfo  {journal} {International journal of solids and structures}\
  }\textbf {\bibinfo {volume} {14}},\ \bibinfo {pages} {161--172} (\bibinfo
  {year} {1978})}\BibitemShut {NoStop}%
\bibitem [{\citenamefont {Sheinman}\ \emph {et~al.}(2012)\citenamefont
  {Sheinman}, \citenamefont {Broedersz},\ and\ \citenamefont
  {MacKintosh}}]{sheinman2012}%
  \BibitemOpen
  \bibfield  {author} {\bibinfo {author} {\bibfnamefont {M}~\bibnamefont
  {Sheinman}}, \bibinfo {author} {\bibfnamefont {CP}~\bibnamefont {Broedersz}},
  \ and\ \bibinfo {author} {\bibfnamefont {FC}~\bibnamefont {MacKintosh}},\
  }\bibfield  {title} {\enquote {\bibinfo {title} {Nonlinear effective-medium
  theory of disordered spring networks},}\ }\href@noop {} {\bibfield  {journal}
  {\bibinfo  {journal} {Physical Review E}\ }\textbf {\bibinfo {volume} {85}},\
  \bibinfo {pages} {021801} (\bibinfo {year} {2012})}\BibitemShut {NoStop}%
\bibitem [{\citenamefont {Merkel}\ \emph {et~al.}(2019)\citenamefont {Merkel},
  \citenamefont {Baumgarten}, \citenamefont {Tighe},\ and\ \citenamefont
  {Manning}}]{merkel2019}%
  \BibitemOpen
  \bibfield  {author} {\bibinfo {author} {\bibfnamefont {Matthias}\
  \bibnamefont {Merkel}}, \bibinfo {author} {\bibfnamefont {Karsten}\
  \bibnamefont {Baumgarten}}, \bibinfo {author} {\bibfnamefont {Brian~P}\
  \bibnamefont {Tighe}}, \ and\ \bibinfo {author} {\bibfnamefont {M~Lisa}\
  \bibnamefont {Manning}},\ }\bibfield  {title} {\enquote {\bibinfo {title} {A
  minimal-length approach unifies rigidity in underconstrained materials},}\
  }\href@noop {} {\bibfield  {journal} {\bibinfo  {journal} {Proceedings of the
  National Academy of Sciences}\ }\textbf {\bibinfo {volume} {116}},\ \bibinfo
  {pages} {6560--6568} (\bibinfo {year} {2019})}\BibitemShut {NoStop}%
\bibitem [{\citenamefont {Vermeulen}\ \emph {et~al.}(2017)\citenamefont
  {Vermeulen}, \citenamefont {Bose}, \citenamefont {Storm},\ and\ \citenamefont
  {Ellenbroek}}]{vermeulen2017}%
  \BibitemOpen
  \bibfield  {author} {\bibinfo {author} {\bibfnamefont {Mathijs~FJ}\
  \bibnamefont {Vermeulen}}, \bibinfo {author} {\bibfnamefont {Anwesha}\
  \bibnamefont {Bose}}, \bibinfo {author} {\bibfnamefont {Cornelis}\
  \bibnamefont {Storm}}, \ and\ \bibinfo {author} {\bibfnamefont {Wouter~G}\
  \bibnamefont {Ellenbroek}},\ }\bibfield  {title} {\enquote {\bibinfo {title}
  {Geometry and the onset of rigidity in a disordered network},}\ }\href@noop
  {} {\bibfield  {journal} {\bibinfo  {journal} {Physical Review E}\ }\textbf
  {\bibinfo {volume} {96}},\ \bibinfo {pages} {053003} (\bibinfo {year}
  {2017})}\BibitemShut {NoStop}%
\bibitem [{\citenamefont {Arzash}\ \emph {et~al.}(2019)\citenamefont {Arzash},
  \citenamefont {Shivers}, \citenamefont {Licup}, \citenamefont {Sharma},\ and\
  \citenamefont {MacKintosh}}]{arzash2019}%
  \BibitemOpen
  \bibfield  {author} {\bibinfo {author} {\bibfnamefont {Sadjad}\ \bibnamefont
  {Arzash}}, \bibinfo {author} {\bibfnamefont {Jordan~L}\ \bibnamefont
  {Shivers}}, \bibinfo {author} {\bibfnamefont {Albert~J}\ \bibnamefont
  {Licup}}, \bibinfo {author} {\bibfnamefont {Abhinav}\ \bibnamefont {Sharma}},
  \ and\ \bibinfo {author} {\bibfnamefont {Fred~C}\ \bibnamefont
  {MacKintosh}},\ }\bibfield  {title} {\enquote {\bibinfo {title}
  {Stress-stabilized subisostatic fiber networks in a ropelike limit},}\
  }\href@noop {} {\bibfield  {journal} {\bibinfo  {journal} {Physical Review
  E}\ }\textbf {\bibinfo {volume} {99}},\ \bibinfo {pages} {042412} (\bibinfo
  {year} {2019})}\BibitemShut {NoStop}%
\bibitem [{\citenamefont {Damavandi}\ \emph
  {et~al.}(2022{\natexlab{a}})\citenamefont {Damavandi}, \citenamefont {Hagh},
  \citenamefont {Santangelo},\ and\ \citenamefont {Manning}}]{damavandi2021}%
  \BibitemOpen
  \bibfield  {author} {\bibinfo {author} {\bibfnamefont {Ojan~Khatib}\
  \bibnamefont {Damavandi}}, \bibinfo {author} {\bibfnamefont {Varda~F}\
  \bibnamefont {Hagh}}, \bibinfo {author} {\bibfnamefont {Christian~D}\
  \bibnamefont {Santangelo}}, \ and\ \bibinfo {author} {\bibfnamefont {M~Lisa}\
  \bibnamefont {Manning}},\ }\bibfield  {title} {\enquote {\bibinfo {title}
  {Energetic rigidity. i. a unifying theory of mechanical stability},}\
  }\href@noop {} {\bibfield  {journal} {\bibinfo  {journal} {Physical Review
  E}\ }\textbf {\bibinfo {volume} {105}},\ \bibinfo {pages} {025003} (\bibinfo
  {year} {2022}{\natexlab{a}})}\BibitemShut {NoStop}%
\bibitem [{\citenamefont {Damavandi}\ \emph
  {et~al.}(2022{\natexlab{b}})\citenamefont {Damavandi}, \citenamefont {Hagh},
  \citenamefont {Santangelo},\ and\ \citenamefont {Manning}}]{damavandi2022}%
  \BibitemOpen
  \bibfield  {author} {\bibinfo {author} {\bibfnamefont {Ojan~Khatib}\
  \bibnamefont {Damavandi}}, \bibinfo {author} {\bibfnamefont {Varda~F}\
  \bibnamefont {Hagh}}, \bibinfo {author} {\bibfnamefont {Christian~D}\
  \bibnamefont {Santangelo}}, \ and\ \bibinfo {author} {\bibfnamefont {M~Lisa}\
  \bibnamefont {Manning}},\ }\bibfield  {title} {\enquote {\bibinfo {title}
  {Energetic rigidity. ii. applications in examples of biological and
  underconstrained materials},}\ }\href@noop {} {\bibfield  {journal} {\bibinfo
   {journal} {Physical Review E}\ }\textbf {\bibinfo {volume} {105}},\ \bibinfo
  {pages} {025004} (\bibinfo {year} {2022}{\natexlab{b}})}\BibitemShut
  {NoStop}%
\bibitem [{\citenamefont {Zhang}\ \emph {et~al.}(2021)\citenamefont {Zhang},
  \citenamefont {Stanifer}, \citenamefont {Vasisht}, \citenamefont {Zhang},
  \citenamefont {Del~Gado},\ and\ \citenamefont {Mao}}]{zhang2021}%
  \BibitemOpen
  \bibfield  {author} {\bibinfo {author} {\bibfnamefont {Shang}\ \bibnamefont
  {Zhang}}, \bibinfo {author} {\bibfnamefont {Ethan}\ \bibnamefont {Stanifer}},
  \bibinfo {author} {\bibfnamefont {Vishwas}\ \bibnamefont {Vasisht}}, \bibinfo
  {author} {\bibfnamefont {Leyou}\ \bibnamefont {Zhang}}, \bibinfo {author}
  {\bibfnamefont {Emanuela}\ \bibnamefont {Del~Gado}}, \ and\ \bibinfo {author}
  {\bibfnamefont {Xiaoming}\ \bibnamefont {Mao}},\ }\bibfield  {title}
  {\enquote {\bibinfo {title} {Prestressed elasticity of amorphous solids},}\
  }\href@noop {} {\bibfield  {journal} {\bibinfo  {journal} {arXiv preprint
  arXiv:2110.07146}\ } (\bibinfo {year} {2021})}\BibitemShut {NoStop}%
\bibitem [{\citenamefont {Lee}\ and\ \citenamefont {Merkel}(2022)}]{lee2022}%
  \BibitemOpen
  \bibfield  {author} {\bibinfo {author} {\bibfnamefont {Cheng-Tai}\
  \bibnamefont {Lee}}\ and\ \bibinfo {author} {\bibfnamefont {Matthias}\
  \bibnamefont {Merkel}},\ }\bibfield  {title} {\enquote {\bibinfo {title}
  {Stiffening of under-constrained spring networks under isotropic strain},}\
  }\href@noop {} {\bibfield  {journal} {\bibinfo  {journal} {arXiv preprint
  arXiv:2201.05385}\ } (\bibinfo {year} {2022})}\BibitemShut {NoStop}%
\bibitem [{\citenamefont {Rudnick}\ and\ \citenamefont
  {Gaspari}(1991)}]{rudnick1991}%
  \BibitemOpen
  \bibfield  {author} {\bibinfo {author} {\bibfnamefont {Joseph}\ \bibnamefont
  {Rudnick}}\ and\ \bibinfo {author} {\bibfnamefont {George}\ \bibnamefont
  {Gaspari}},\ }\bibfield  {title} {\enquote {\bibinfo {title} {The shapes and
  sizes of closed, pressurized random walks},}\ }\href@noop {} {\bibfield
  {journal} {\bibinfo  {journal} {Science}\ }\textbf {\bibinfo {volume}
  {252}},\ \bibinfo {pages} {422--424} (\bibinfo {year} {1991})}\BibitemShut
  {NoStop}%
\bibitem [{\citenamefont {Leibler}\ \emph {et~al.}(1987)\citenamefont
  {Leibler}, \citenamefont {Singh},\ and\ \citenamefont
  {Fisher}}]{leibler1987}%
  \BibitemOpen
  \bibfield  {author} {\bibinfo {author} {\bibfnamefont {Stanislas}\
  \bibnamefont {Leibler}}, \bibinfo {author} {\bibfnamefont {Rajiv~RP}\
  \bibnamefont {Singh}}, \ and\ \bibinfo {author} {\bibfnamefont {Michael~E}\
  \bibnamefont {Fisher}},\ }\bibfield  {title} {\enquote {\bibinfo {title}
  {Thermodynamic behavior of two-dimensional vesicles},}\ }\href@noop {}
  {\bibfield  {journal} {\bibinfo  {journal} {Physical review letters}\
  }\textbf {\bibinfo {volume} {59}},\ \bibinfo {pages} {1989} (\bibinfo {year}
  {1987})}\BibitemShut {NoStop}%
\bibitem [{\citenamefont {Haleva}\ and\ \citenamefont
  {Diamant}(2006)}]{haleva2006}%
  \BibitemOpen
  \bibfield  {author} {\bibinfo {author} {\bibfnamefont {Emir}\ \bibnamefont
  {Haleva}}\ and\ \bibinfo {author} {\bibfnamefont {Haim}\ \bibnamefont
  {Diamant}},\ }\bibfield  {title} {\enquote {\bibinfo {title} {Smoothening
  transition of a two-dimensional pressurized polymer ring},}\ }\href@noop {}
  {\bibfield  {journal} {\bibinfo  {journal} {The European Physical Journal E}\
  }\textbf {\bibinfo {volume} {19}},\ \bibinfo {pages} {461--469} (\bibinfo
  {year} {2006})}\BibitemShut {NoStop}%
\bibitem [{\citenamefont {Mitra}\ \emph {et~al.}(2012)\citenamefont {Mitra},
  \citenamefont {Menon},\ and\ \citenamefont {Rajesh}}]{mitra2012}%
  \BibitemOpen
  \bibfield  {author} {\bibinfo {author} {\bibfnamefont {Mithun~K}\
  \bibnamefont {Mitra}}, \bibinfo {author} {\bibfnamefont {Gautam~I}\
  \bibnamefont {Menon}}, \ and\ \bibinfo {author} {\bibfnamefont
  {R}~\bibnamefont {Rajesh}},\ }\bibfield  {title} {\enquote {\bibinfo {title}
  {Thermodynamic behaviour of two-dimensional vesicles revisited},}\
  }\href@noop {} {\bibfield  {journal} {\bibinfo  {journal} {The European
  Physical Journal E}\ }\textbf {\bibinfo {volume} {35}},\ \bibinfo {pages}
  {1--8} (\bibinfo {year} {2012})}\BibitemShut {NoStop}%
\bibitem [{Note1()}]{Note1}%
  \BibitemOpen
  \bibinfo {note} {There are more possible configurations for a crumpled
  polygon as compared to the inflated configurations.}\BibitemShut {Stop}%
\bibitem [{\citenamefont {Gaspari}\ \emph {et~al.}(1993)\citenamefont
  {Gaspari}, \citenamefont {Rudnick},\ and\ \citenamefont
  {Beldjenna}}]{gaspari1993}%
  \BibitemOpen
  \bibfield  {author} {\bibinfo {author} {\bibfnamefont {George}\ \bibnamefont
  {Gaspari}}, \bibinfo {author} {\bibfnamefont {Joseph}\ \bibnamefont
  {Rudnick}}, \ and\ \bibinfo {author} {\bibfnamefont {Arezki}\ \bibnamefont
  {Beldjenna}},\ }\bibfield  {title} {\enquote {\bibinfo {title} {The shapes
  and sizes of two-dimensional pressurized self-intersecting rings, as models
  for two-dimensional vesicles},}\ }\href@noop {} {\bibfield  {journal}
  {\bibinfo  {journal} {Journal of Physics A: Mathematical and General}\
  }\textbf {\bibinfo {volume} {26}},\ \bibinfo {pages} {1} (\bibinfo {year}
  {1993})}\BibitemShut {NoStop}%
\bibitem [{\citenamefont {Cauchy}(1813)}]{cauchy1813}%
  \BibitemOpen
  \bibfield  {author} {\bibinfo {author} {\bibfnamefont {Augustin~Louis}\
  \bibnamefont {Cauchy}},\ }\bibfield  {title} {\enquote {\bibinfo {title} {Sur
  les polygones et les polyedres, seconde m{\'e}moire},}\ }\href@noop {}
  {\bibfield  {journal} {\bibinfo  {journal} {J. Ecole Polytechnique}\ }\textbf
  {\bibinfo {volume} {9}},\ \bibinfo {pages} {87--98} (\bibinfo {year}
  {1813})}\BibitemShut {NoStop}%
\bibitem [{\citenamefont {Aigner}\ \emph {et~al.}(2010)\citenamefont {Aigner},
  \citenamefont {Ziegler}, \citenamefont {Hofmann},\ and\ \citenamefont
  {Erdos}}]{aigner2010}%
  \BibitemOpen
  \bibfield  {author} {\bibinfo {author} {\bibfnamefont {Martin}\ \bibnamefont
  {Aigner}}, \bibinfo {author} {\bibfnamefont {G{\"u}nter~M}\ \bibnamefont
  {Ziegler}}, \bibinfo {author} {\bibfnamefont {Karl~H}\ \bibnamefont
  {Hofmann}}, \ and\ \bibinfo {author} {\bibfnamefont {Paul}\ \bibnamefont
  {Erdos}},\ }\href@noop {} {\emph {\bibinfo {title} {Proofs from the Book}}},\
  Vol.\ \bibinfo {volume} {274}\ (\bibinfo  {publisher} {Springer},\ \bibinfo
  {year} {2010})\BibitemShut {NoStop}%
\bibitem [{\citenamefont {Pak}(2010)}]{pak2010lectures}%
  \BibitemOpen
  \bibfield  {author} {\bibinfo {author} {\bibfnamefont {Igor}\ \bibnamefont
  {Pak}},\ }\bibfield  {title} {\enquote {\bibinfo {title} {Lectures on
  discrete and polyhedral geometry},}\ }\href@noop {} {\  (\bibinfo {year}
  {2010})}\BibitemShut {NoStop}%
\bibitem [{\citenamefont {O'Rourke}(1994)}]{rourke}%
  \BibitemOpen
  \bibfield  {author} {\bibinfo {author} {\bibfnamefont {Joseph}\ \bibnamefont
  {O'Rourke}},\ }\href@noop {} {\emph {\bibinfo {title} {Computational Geometry
  in C}}}\ (\bibinfo  {publisher} {Cambridge University Press},\ \bibinfo
  {year} {1994})\BibitemShut {NoStop}%
\bibitem [{Note2()}]{Note2}%
  \BibitemOpen
  \bibinfo {note} {Graham's algorithm picks a point within the convex hull of
  these set of points. We are assuming here that the `center of mass' of the
  points is within the convex hull of the points. We perform this approximation
  for numerical simplicity.}\BibitemShut {Stop}%
\bibitem [{\citenamefont {Galassi}\ \emph {et~al.}(2002)\citenamefont
  {Galassi}, \citenamefont {Davies}, \citenamefont {Theiler}, \citenamefont
  {Gough}, \citenamefont {Jungman}, \citenamefont {Alken}, \citenamefont
  {Booth}, \citenamefont {Rossi},\ and\ \citenamefont
  {Ulerich}}]{galassi2002gnu}%
  \BibitemOpen
  \bibfield  {author} {\bibinfo {author} {\bibfnamefont {Mark}\ \bibnamefont
  {Galassi}}, \bibinfo {author} {\bibfnamefont {Jim}\ \bibnamefont {Davies}},
  \bibinfo {author} {\bibfnamefont {James}\ \bibnamefont {Theiler}}, \bibinfo
  {author} {\bibfnamefont {Brian}\ \bibnamefont {Gough}}, \bibinfo {author}
  {\bibfnamefont {Gerard}\ \bibnamefont {Jungman}}, \bibinfo {author}
  {\bibfnamefont {Patrick}\ \bibnamefont {Alken}}, \bibinfo {author}
  {\bibfnamefont {Michael}\ \bibnamefont {Booth}}, \bibinfo {author}
  {\bibfnamefont {Fabrice}\ \bibnamefont {Rossi}}, \ and\ \bibinfo {author}
  {\bibfnamefont {Rhys}\ \bibnamefont {Ulerich}},\ }\href@noop {} {\emph
  {\bibinfo {title} {GNU scientific library}}}\ (\bibinfo  {publisher} {Network
  Theory Limited},\ \bibinfo {year} {2002})\BibitemShut {NoStop}%
\bibitem [{\citenamefont {Devadoss}\ and\ \citenamefont
  {O'Rourke}(2011)}]{devadoss2011}%
  \BibitemOpen
  \bibfield  {author} {\bibinfo {author} {\bibfnamefont {Satyan~L}\
  \bibnamefont {Devadoss}}\ and\ \bibinfo {author} {\bibfnamefont {Joseph}\
  \bibnamefont {O'Rourke}},\ }\href@noop {} {\emph {\bibinfo {title} {Discrete
  and computational geometry}}}\ (\bibinfo  {publisher} {Princeton University
  Press},\ \bibinfo {year} {2011})\BibitemShut {NoStop}%
\bibitem [{Note3()}]{Note3}%
  \BibitemOpen
  \bibinfo {note} {If the increase in $A_0$ is small compared to the area
  enclosed by the rhombus formed by the dashed and solid lines, then the
  discrete flip of the edges in Fig. (\ref {flip}a) will be an
  over-compensation to the imposed strain and incurs an energy cost. Here, the
  response of the polygon will be less local. The nearest neighbor and the
  next-to-nearest neighbor edges of the mouth will continuously open to
  accommodate small increases in $A_0$.}\BibitemShut {Stop}%
\bibitem [{\citenamefont {Niven}(1981)}]{niven1981}%
  \BibitemOpen
  \bibfield  {author} {\bibinfo {author} {\bibfnamefont {Ivan}\ \bibnamefont
  {Niven}},\ }\href@noop {} {\emph {\bibinfo {title} {Maxima and minima without
  calculus}}},\ \bibinfo {number} {6}\ (\bibinfo  {publisher} {Cambridge
  University Press},\ \bibinfo {year} {1981})\BibitemShut {NoStop}%
\bibitem [{\citenamefont {Pinelis}(2005)}]{pinelis2005}%
  \BibitemOpen
  \bibfield  {author} {\bibinfo {author} {\bibfnamefont {Iosif}\ \bibnamefont
  {Pinelis}},\ }\bibfield  {title} {\enquote {\bibinfo {title} {Cyclic polygons
  with given edge lengths: existence and uniqueness},}\ }\href@noop {}
  {\bibfield  {journal} {\bibinfo  {journal} {Journal of Geometry}\ }\textbf
  {\bibinfo {volume} {82}},\ \bibinfo {pages} {156--171} (\bibinfo {year}
  {2005})}\BibitemShut {NoStop}%
\bibitem [{\citenamefont
  {Alexander}(1998{\natexlab{b}})}]{alexander1998amorphous}%
  \BibitemOpen
  \bibfield  {author} {\bibinfo {author} {\bibfnamefont {Shlomo}\ \bibnamefont
  {Alexander}},\ }\bibfield  {title} {\enquote {\bibinfo {title} {Amorphous
  solids: their structure, lattice dynamics and elasticity},}\ }\href@noop {}
  {\bibfield  {journal} {\bibinfo  {journal} {Physics reports}\ }\textbf
  {\bibinfo {volume} {296}},\ \bibinfo {pages} {65--236} (\bibinfo {year}
  {1998}{\natexlab{b}})}\BibitemShut {NoStop}%
\bibitem [{\citenamefont {Meurer}\ \emph {et~al.}(2017)\citenamefont {Meurer},
  \citenamefont {Smith}, \citenamefont {Paprocki}, \citenamefont
  {{\v{C}}ert{\'\i}k}, \citenamefont {Kirpichev}, \citenamefont {Rocklin},
  \citenamefont {Kumar}, \citenamefont {Ivanov}, \citenamefont {Moore},
  \citenamefont {Singh} \emph {et~al.}}]{meurer2017sympy}%
  \BibitemOpen
  \bibfield  {author} {\bibinfo {author} {\bibfnamefont {Aaron}\ \bibnamefont
  {Meurer}}, \bibinfo {author} {\bibfnamefont {Christopher~P}\ \bibnamefont
  {Smith}}, \bibinfo {author} {\bibfnamefont {Mateusz}\ \bibnamefont
  {Paprocki}}, \bibinfo {author} {\bibfnamefont {Ond{\v{r}}ej}\ \bibnamefont
  {{\v{C}}ert{\'\i}k}}, \bibinfo {author} {\bibfnamefont {Sergey~B}\
  \bibnamefont {Kirpichev}}, \bibinfo {author} {\bibfnamefont {Matthew}\
  \bibnamefont {Rocklin}}, \bibinfo {author} {\bibfnamefont {AMiT}\
  \bibnamefont {Kumar}}, \bibinfo {author} {\bibfnamefont {Sergiu}\
  \bibnamefont {Ivanov}}, \bibinfo {author} {\bibfnamefont {Jason~K}\
  \bibnamefont {Moore}}, \bibinfo {author} {\bibfnamefont {Sartaj}\
  \bibnamefont {Singh}},  \emph {et~al.},\ }\bibfield  {title} {\enquote
  {\bibinfo {title} {Sympy: symbolic computing in python},}\ }\href@noop {}
  {\bibfield  {journal} {\bibinfo  {journal} {PeerJ Computer Science}\ }\textbf
  {\bibinfo {volume} {3}},\ \bibinfo {pages} {e103} (\bibinfo {year}
  {2017})}\BibitemShut {NoStop}%
\bibitem [{\citenamefont {Robbins}(1994)}]{robbins1994}%
  \BibitemOpen
  \bibfield  {author} {\bibinfo {author} {\bibfnamefont {David~P}\ \bibnamefont
  {Robbins}},\ }\bibfield  {title} {\enquote {\bibinfo {title} {Areas of
  polygons inscribed in a circle},}\ }\href@noop {} {\bibfield  {journal}
  {\bibinfo  {journal} {Discrete \& Computational Geometry}\ }\textbf {\bibinfo
  {volume} {12}},\ \bibinfo {pages} {223--236} (\bibinfo {year}
  {1994})}\BibitemShut {NoStop}%
\bibitem [{\citenamefont {Robbins}(1995)}]{robbins1995}%
  \BibitemOpen
  \bibfield  {author} {\bibinfo {author} {\bibfnamefont {David~P}\ \bibnamefont
  {Robbins}},\ }\bibfield  {title} {\enquote {\bibinfo {title} {Areas of
  polygons inscribed in a circle},}\ }\href@noop {} {\bibfield  {journal}
  {\bibinfo  {journal} {The American mathematical monthly}\ }\textbf {\bibinfo
  {volume} {102}},\ \bibinfo {pages} {523--530} (\bibinfo {year}
  {1995})}\BibitemShut {NoStop}%
\bibitem [{\citenamefont {Sussman}(2017)}]{sussman2017}%
  \BibitemOpen
  \bibfield  {author} {\bibinfo {author} {\bibfnamefont {Daniel~M}\
  \bibnamefont {Sussman}},\ }\bibfield  {title} {\enquote {\bibinfo {title}
  {cellgpu: Massively parallel simulations of dynamic vertex models},}\
  }\href@noop {} {\bibfield  {journal} {\bibinfo  {journal} {Computer Physics
  Communications}\ }\textbf {\bibinfo {volume} {219}},\ \bibinfo {pages}
  {400--406} (\bibinfo {year} {2017})}\BibitemShut {NoStop}%
\bibitem [{\citenamefont {Guiducci}\ \emph {et~al.}(2015)\citenamefont
  {Guiducci}, \citenamefont {Weaver}, \citenamefont {Br{\'e}chet},
  \citenamefont {Fratzl},\ and\ \citenamefont {Dunlop}}]{guiducci2015}%
  \BibitemOpen
  \bibfield  {author} {\bibinfo {author} {\bibfnamefont {Lorenzo}\ \bibnamefont
  {Guiducci}}, \bibinfo {author} {\bibfnamefont {James~C}\ \bibnamefont
  {Weaver}}, \bibinfo {author} {\bibfnamefont {Yves~JM}\ \bibnamefont
  {Br{\'e}chet}}, \bibinfo {author} {\bibfnamefont {Peter}\ \bibnamefont
  {Fratzl}}, \ and\ \bibinfo {author} {\bibfnamefont {John~WC}\ \bibnamefont
  {Dunlop}},\ }\bibfield  {title} {\enquote {\bibinfo {title} {The geometric
  design and fabrication of actuating cellular structures},}\ }\href@noop {}
  {\bibfield  {journal} {\bibinfo  {journal} {Advanced Materials Interfaces}\
  }\textbf {\bibinfo {volume} {2}},\ \bibinfo {pages} {1500011} (\bibinfo
  {year} {2015})}\BibitemShut {NoStop}%
\bibitem [{\citenamefont {Connelly}\ \emph {et~al.}(2000)\citenamefont
  {Connelly}, \citenamefont {Demaine},\ and\ \citenamefont
  {Rote}}]{connelly2000straightening}%
  \BibitemOpen
  \bibfield  {author} {\bibinfo {author} {\bibfnamefont {Robert}\ \bibnamefont
  {Connelly}}, \bibinfo {author} {\bibfnamefont {Erik~D}\ \bibnamefont
  {Demaine}}, \ and\ \bibinfo {author} {\bibfnamefont {G{\"u}nter}\
  \bibnamefont {Rote}},\ }\bibfield  {title} {\enquote {\bibinfo {title}
  {Straightening polygonal arcs and convexifying polygonal cycles},}\ }in\
  \href@noop {} {\emph {\bibinfo {booktitle} {Proceedings 41st Annual Symposium
  on Foundations of Computer Science}}}\ (\bibinfo {organization} {IEEE},\
  \bibinfo {year} {2000})\ pp.\ \bibinfo {pages} {432--442}\BibitemShut
  {NoStop}%
\bibitem [{Note4()}]{Note4}%
  \BibitemOpen
  \bibinfo {note} {Struts are elastic rods which are in compression and hence
  want to expand.}\BibitemShut {Stop}%
\bibitem [{\citenamefont {Cantarella}\ \emph {et~al.}(2004)\citenamefont
  {Cantarella}, \citenamefont {Demaine}, \citenamefont {Iben},\ and\
  \citenamefont {O'Brien}}]{cantarella2004}%
  \BibitemOpen
  \bibfield  {author} {\bibinfo {author} {\bibfnamefont {Jason~H}\ \bibnamefont
  {Cantarella}}, \bibinfo {author} {\bibfnamefont {Erik~D}\ \bibnamefont
  {Demaine}}, \bibinfo {author} {\bibfnamefont {Hayley~N}\ \bibnamefont
  {Iben}}, \ and\ \bibinfo {author} {\bibfnamefont {James~F}\ \bibnamefont
  {O'Brien}},\ }\bibfield  {title} {\enquote {\bibinfo {title} {An
  energy-driven approach to linkage unfolding},}\ }in\ \href@noop {} {\emph
  {\bibinfo {booktitle} {Proceedings of the twentieth annual symposium on
  Computational geometry}}}\ (\bibinfo {year} {2004})\ pp.\ \bibinfo {pages}
  {134--143}\BibitemShut {NoStop}%
\bibitem [{\citenamefont {Mueth}\ \emph {et~al.}(1998)\citenamefont {Mueth},
  \citenamefont {Jaeger},\ and\ \citenamefont {Nagel}}]{Mueth1998}%
  \BibitemOpen
  \bibfield  {author} {\bibinfo {author} {\bibfnamefont {Daniel~M.}\
  \bibnamefont {Mueth}}, \bibinfo {author} {\bibfnamefont {Heinrich~M.}\
  \bibnamefont {Jaeger}}, \ and\ \bibinfo {author} {\bibfnamefont {Sidney~R.}\
  \bibnamefont {Nagel}},\ }\bibfield  {title} {\enquote {\bibinfo {title}
  {Force distribution in a granular medium},}\ }\href@noop {} {\bibfield
  {journal} {\bibinfo  {journal} {Physical Review E}\ }\textbf {\bibinfo
  {volume} {57}},\ \bibinfo {pages} {3164} (\bibinfo {year}
  {1998})}\BibitemShut {NoStop}%
\bibitem [{\citenamefont {Majmudar}\ and\ \citenamefont
  {Behringer}(2005)}]{Majmudar2005}%
  \BibitemOpen
  \bibfield  {author} {\bibinfo {author} {\bibfnamefont {Trushant~S.}\
  \bibnamefont {Majmudar}}\ and\ \bibinfo {author} {\bibfnamefont {Robert~P.}\
  \bibnamefont {Behringer}},\ }\bibfield  {title} {\enquote {\bibinfo {title}
  {Contact force measurements and stress-induced anisotropy in granular
  materials},}\ }\href@noop {} {\bibfield  {journal} {\bibinfo  {journal}
  {Nature}\ }\textbf {\bibinfo {volume} {435}},\ \bibinfo {pages} {1079--1082}
  (\bibinfo {year} {2005})}\BibitemShut {NoStop}%
\bibitem [{\citenamefont {Sarkar}\ \emph {et~al.}(2013)\citenamefont {Sarkar},
  \citenamefont {Bi}, \citenamefont {Zhang}, \citenamefont {Behringer},\ and\
  \citenamefont {Chakraborty}}]{Sarkar2013}%
  \BibitemOpen
  \bibfield  {author} {\bibinfo {author} {\bibfnamefont {Sumantra}\
  \bibnamefont {Sarkar}}, \bibinfo {author} {\bibfnamefont {Dapeng}\
  \bibnamefont {Bi}}, \bibinfo {author} {\bibfnamefont {Jie}\ \bibnamefont
  {Zhang}}, \bibinfo {author} {\bibfnamefont {R.~P.}\ \bibnamefont
  {Behringer}}, \ and\ \bibinfo {author} {\bibfnamefont {Bulbul}\ \bibnamefont
  {Chakraborty}},\ }\bibfield  {title} {\enquote {\bibinfo {title} {Origin of
  rigidity in dry granular solids},}\ }\href@noop {} {\bibfield  {journal}
  {\bibinfo  {journal} {Physical review letters}\ }\textbf {\bibinfo {volume}
  {111}},\ \bibinfo {pages} {068301} (\bibinfo {year} {2013})}\BibitemShut
  {NoStop}%
\end{thebibliography}%
\end{document}